\documentclass[useAMS,usenatbib]{mn2e}

\usepackage{epsfig}
\usepackage{myaasmacros}
\usepackage{amsmath}
\usepackage{amsfonts}
\usepackage{subfig}
\usepackage{comment}
\usepackage{color}

\title[Influence of the environmental history] 
{The influence of the environmental history on quenching star formation
  in a $\Lambda$CDM
  universe}  
\author[Hirschmann et al.]{Michaela Hirschmann$^{1,2}$\thanks{E-mail:
hirschma@iap.fr}, Gabriella De Lucia$^{1}$, Dave Wilman$^{3}$,
Simone Weinmann$^{4}$, \newauthor Angela Iovino$^5$, Olga
Cucciati$^{6}$, Stefano Zibetti$^7$, \'Alvaro Villalobos$^1$\\
$^{1}$INAF - Astronomical Observatory of Trieste, via G.B. Tiepolo 11,
I-34143 Trieste, Italy\\
$^{2}$UPMC-CNRS, UMR7095, Institut d' Astrophysique de Paris, Boulevard
Arago, F-75014
Paris, France\\ 
$^3$Max-Planck-Institute for Extraterrestrial Physics,
Giessenbachstrasse, D-85748 Garching, Germany\\
$^4$ Leiden Observatory, Leiden University, PO Box 9513, 2300 RA
Leiden, the Netherlands\\
$^5$INAF - Astronomical Observatory of Brera, via Brera 28, I-20159
Milano, Italy\\
$^6$INAF - Astronomical Observatory of Bologna, via Ranzani 1, I-40127
Bologna, Italy\\
$^7$INAF - Astrophysical Observatory of Arcetri, Largo Enrico Fermi 5,
I-501125 Firenze, Italy}

\begin{document}

\date{Accepted ???. Received ??? in original form ???}

\pagerange{\pageref{firstpage}--\pageref{lastpage}} \pubyear{2002}

\maketitle

\label{firstpage}

\begin{abstract}
We present a detailed analysis of the influence of the environment and 
of the environmental history on quenching star formation in central
and satellite galaxies in the local Universe. We take advantage of
publicly available galaxy catalogues obtained from applying a galaxy
formation model to the Millennium simulation. In addition to halo
mass, we consider the local density of galaxies within various fixed
scales. Comparing our model predictions to observational data (SDSS),
we demonstrate that the models are failing to reproduce the observed
density dependence of the quiescent galaxy fraction in several
aspects: for most of the stellar mass ranges and densities explored,
models cannot reproduce the observed similar behaviour of centrals and
satellites, they slightly under-estimate the quiescent fraction of
centrals and significantly over-estimate that of satellites. We show
that in the models, the density dependence of the quiescent central
galaxies is caused by a fraction of ``backsplash" centrals which have
been satellites in the past. The observed stronger density dependence on
scales of $0.2-1$~Mpc may, however, indicate additional environmental
processes working on central galaxies. Turning to satellite galaxies,
the density dependence of their quiescent fractions reflects a 
dependence on the time spent orbiting within a parent halo,
correlating strongly with halo mass and distance from the halo
centre. Comparisons with observational estimates suggest relatively
long gas consumption time scales of roughly 5~Gyr in low mass
satellite galaxies. The quenching time scales decrease with increasing
satellite stellar mass. Overall, a change in modelling both internal
processes and environmental processes is required for improving
currently used galaxy formation models. 
\end{abstract}

\begin{keywords}
galaxies: evolution - galaxies: formation
\end{keywords}

%*****************************************************************************************************
%*****************************************************************************************************
\section{Introduction}\label{intro}
%*****************************************************************************************************
%*****************************************************************************************************

%Basic, well-known observational facts with respect to environment
It has long been known that the properties of galaxies are
strongly dependent on the environment in which they are
residing. Observations have revealed that red, early-type galaxies are
preferentially located in high-density environments, while blue,
late-type galaxies dominate the galaxy population in low-density
environments (\citealp{Oemler74, Davis76, Dressler80, Balogh97,
  Poggianti99}). In the last decade, observational studies based on
large spectroscopic and photometric surveys have invested significant
effort to understand the role of environment on galaxy formation and
evolution (e.g. \citealp{Balogh04, Kauffmann04} and
\citealp{Blanton09} for a recent review). It remains, however, heavily
debated to what extent the properties of galaxies are determined by
external processes (nurture, i.e. interaction with  other galaxies or
with the local environment) or internal processes (nature, e.g. gas
cooling, star formation, feedback processes etc.). 

%Describe the nurture scenario for satellites
In the nurture scenario, different environmental processes are thought
to be effective in suppressing star formation and altering the
morphology of a galaxy living in a ``dense'' environment, such as
strangulation (\citealp{Larson80}), ram pressure stripping
(\citealp{Gunn72, Abadi99}), tidal stripping (\citealp{Dekel03,
  Diemand07}), or harassment (\citealp{Farouki81, Moore98}).  All
these processes are expected to primarily affect \textit{satellite}
galaxies (i.e. galaxies orbiting within a halo containing a more
  dominant, {\it central} galaxy), and to have a stronger influence on
  those satellites residing in denser environments. However, their
  relative importance to quench star formation, as a function of
  density or halo mass, remains unclear.  

%Nurture scenario for centrals
Environmental effects might extend to central galaxies beyond the
virial radius of a massive neighbour halo. Some observational studies
(e.g. \citealp{Balogh00, Haines09, vonderLinden10, Rasmussen12,
  Geha12}) find an enhanced quiescent fraction of centrals in the
vicinity of massive neighbour haloes out to four times the virial
radius. Other studies find, however, no such trend. Environmental
trends extending beyond the virial radius might - at least partly - be
due to the fact that such galaxies have been residing inside a larger
structure in the past. \citet{Balogh00} first showed that in N-body
simulations there are particles that have resided within the virial
radius of clusters but have moved outside the virial radius later on. 
Present-day central galaxies could have behaved similarly as shown by
\citet{Mamon04} analytically or by e.g. \citet{Ludlow09},
\citet{Knebe11} and \citet{Bahe13} in numerical simulations. However,
statistically, it remains unclear how strongly central galaxies are
influenced by environmental effects. Thus, whether or not central
galaxy properties depend on the environment on super-halo scales is
currently heavily debated (see e.g. \citealp{Blanton07, Wilman10,
  Bahe13, Wetzel14, Kovac14, Peng12, Woo13}).   
 
%Modelling environment in SAMs
The effect of environmental processes has been theoretically studied
by using semi-empirical models (\citealp{Wetzel12}), numerical
simulations (see \citealp{DeLucia11} for a review) and semi-analytic
models (\citealp{DeLucia07, Kang08, Font08, Kimm09, Weinmann10,
  Guo11}). Most of the currently used galaxy formation models do
not explicitly include environmental processes beyond the virial
radius of haloes. Environmental processes (working on satellites) are
typically described by simplified recipes, e.g. an instantaneous
strangulation of the hot halo gas of satellites. This leads to a
significant over-estimation of the quiescent fraction of model
satellite galaxies compared to observations (e.g. \citealp{Kimm09}).
Some progress has been made by adopting a more gradual stripping of
the hot gas associated with infalling galaxies (e.g. \citealp{Kang08,
  Font08, Weinmann10, Guo11}). The improved models are, however, still
inconsistent with observations (e.g. with regard to the bi-modality of
the star formation rate distributions) indicating the need of further
fundamental changes and modifications.  

%The importance of the environmental history
In the currently favoured hierarchical structure formation scenarios,
based on the Cold Dark Matter (CDM) model (\citealp{Peebles65,
  White78, Blumenthal85}), locally over-dense regions collapse and
form virialised dark matter haloes. Small objects form first and
subsequently grow into larger systems via merging and smooth particle
accretion. This implies that with evolving cosmic time, galaxies join
more and more massive systems and therefore, experience a large
variety of environments during their life time. In this context,
external and internal effects are strongly connected and heavily
intertwined, and disentangling the influence of nature and nurture is
difficult.    

A different evolution of galaxies in different environments is
expected to leave an imprint on the observable galaxy
properties. Haloes in over-dense regions on average form earlier 
and merge more rapidly than those residing in regions of average
density. In this respect, a detailed quantification of the influence of
\textit{the environmental history} on galaxy properties is still
lacking, and appears to be of crucial importance in order to correctly
explain and understand the observed environmental trends.  A few
recent studies have started to explore these issues such as
\citet{Berrier09, McGee09, DeLucia12, Hirschmann13} and
\citet{Wetzel13, Wetzel14}. \citet{Berrier09} using numerical
simulations and \citet{McGee09} using semi-analytic models have
studied the accretion history of galaxies onto clusters to investigate
and to quantify the importance of pre-processing in galaxy groups.
\citet{Hirschmann13} have investigated the environmental history of
isolated galaxies in galaxy formation models and confirm that those
are hardly affected by nurture and environmental effects during their
life time.  

Two recent studies of \citet{Wetzel13, Wetzel14} connect observational
data with merger trees extracted from N-body simulations and adopt
simple parametrisations for the evolution of star formation in
satellites before and while their star formation is quenched. They
predict long quenching time-scales for satellites and ejected centrals
(up to 5~Gyrs) increasing with decreasing stellar mass. Finally, a recent
study by \citet{DeLucia12} has focused on the environmental history of
group and cluster galaxies. By comparing galaxy formation models to
observational data, they also predict relatively long quenching time
scales of $5-7\ \mathrm{Gyrs}$ within haloes more massive than $10^{13}
M_{\odot}$, independent of the galaxy stellar mass. Their study
treated the evolution as a function of estimated halo and stellar
masses, but did not consider more detailed quantification of
environment, such as the density of neighbouring galaxies on different
scales. This partly motivates this work.

We present a detailed analysis of the effect of the environmental
history by adopting similar approaches as in \citet{DeLucia12}. We
take advantage of publicly available galaxy merger trees, obtained by
applying the semi-analytic model of \citet{Guo11} to the large
cosmological dark matter Millennium simulation. The merger trees are
analysed in order to study the history of the environment that
galaxies have experienced during their life time, as a function of
environmental density, stellar mass and galaxy type. 
In addition to estimated halo mass, we consider different density
estimators that are commonly used in observational studies. For a
comparison between model predictions and observations, we have
computed densities based on the SDSS DR8 database
(\citealp{Wilman10}). In our analysis, we will particularly focus on
the following points:     
\begin{enumerate}
\item[{\bf 1.}] \noindent First, we will specify the deficiencies of
  current galaxy formation models in predicting the observed
  environmental dependence of quiescent central and satellite
  galaxies. 
\item[{\bf 2.}] \noindent Second, we will explore the physical origin
  of the density dependence of quiescent central and satellite
  galaxies by considering their environmental history.  
\item[{\bf 3.}] \noindent Third, we will discuss possible improvements
  for semi-analytic models by investigating the relevance of
  environmental processes on super-halo scales for quenching central
  galaxies and by constraining star formation quenching timescales for
  satellite galaxies and their dependence on density and stellar mass.
\end{enumerate}
The structure of the paper is as follows. In Section \ref{theory}, we
will briefly introduce the theoretical models and the observational
data. In Section \ref{stellardep} we compare the stellar mass
dependence of the quiescent fraction in observations and models. 
Section \ref{densdep} presents a careful analysis of the density
dependence of quiescent central and satellite galaxies and discusses
the origin of these trends. In Section \ref{envhist} we constrain
quenching time scales for satellite galaxies by considering their
environmental history and comparing it with observational
estimates. Finally, Section \ref{summary} gives a discussion and
summary of this work.\\

%*****************************************************************************************************
%*****************************************************************************************************
\section{Method}\label{theory}
%*****************************************************************************************************
%*****************************************************************************************************

%*****************************************************************************************************
\subsection{Observational data}\label{observations}
%*****************************************************************************************************

\newcommand{\red}{\textcolor{red}}

We use observational data from the SDSS (DR8) database which we
cross-correlate with a modified catalogue of \citet{Wilman10}. This
catalogue lists the neighbours of each galaxy within cylinders of
different radii. Luminosities are computed using SDSS r-band Petrosian 
magnitudes, our adopted cosmology and k-corrected using the {\sc
  k-correct} idl tool \citep{BlantonRoweis07}. We select ``primary''
galaxies with a SDSS r-band magnitude down to $M_r < -18$ at $z=0$,
but as ``neighbour'' galaxies (galaxies used to calculate the density
around a ``primary'' galaxy) we take only galaxies brighter than $M_r
< -20$. The galaxy sample is restricted to redshifts between
$0.015<z<0.08$. This implies a volume-limited, and thus complete,
``neighbour'' sample when assuming the magnitude cut of $M_r <
-20$. Instead, the ``primary'' galaxy sample with $M_r < -18$ is not
volume-limited over the full redshift range. Therefore, we limit
our primary sample to smaller volumes for lower luminosity galaxies,
and correct for volume incompleteness ($V_{\mathrm{max}}$ correction)
throughout this study. 

As a measure for environment, we consider the local densities of our
``primary'' galaxies by calculating the projected surface density of
neighbour galaxies, $\Sigma_{r}$, within a cylinder of depth $\pm
1000~kms^{-1}$ and with a projected radius centred on each primary
galaxy, and selected to trace the environment on a scale of our
choice. In this paper, we consider densities on two different scales,
1~Mpc and 0.2~Mpc. 
The former traces density on the scale of the most massive haloes at
$z=0$ while the latter examines the role of density on much smaller
scales, but still large enough to avoid under-sampling fairly dense
environments with our luminosity limit. 
The projected surface density is then: 
\begin{equation}
\label{dens1mpc}
\Sigma_{r} = N_{r}/(\pi r^2),
\end{equation}
where $r$ is the radius of the cylinder and $N_{r}$ is the number
of neighbours corrected for spectroscopic completeness within this
cylinder. Spectroscopic completeness of neighbours is evaluated and
accounted for as described by \citet{Wilman10}. Within SDSS,
correction factors are typically small, with a minority of large
values in dense regions caused by fibre collisions. To ensure
spectroscopic completeness corrections are feasible on scales up to
3~Mpc we require high ($>99.35\%$, see \citet{Wilman10}) photometric 
completeness within this distance. This provides a total primary
sample of 300,000 galaxies. 

In order to additionally obtain specific star formation rates,
stellar masses, halo masses and the central/satellite status
for the observed galaxies, we have cross-correlated this
catalogue with the corresponding catalogues from
JHU-MPA\footnote{http://www.mpa-garching.mpg.de/SDSS/DR7/} 
and the updated DR7 version of the group catalogue of
\citet{Yang07}. The JHU-MPA DR7 catalogue is also updated with respect
to the published DR4 version as described by 
\citet{Brinchmann04} (SFR) and \citet{Kauffmann03} (stellar
masses). These new catalogues contain not only a larger number of
galaxies, but are also based on an improved methodology for the
estimation of physical properties as described by \citet{Salim07}.
SED fits to the photometry are used to improve stellar masses,
aperture corrections to spectroscopically derived SFRs, and the SFRs
of galaxies where emission lines cannot be used. The group (or halo)
catalogue employs an adaptive friends-of-friends methodology with halo
masses estimated using abundance matching and considering all galaxies 
brighter than an evolution and k-corrected r-band absolute magnitude
of $-19.5$ (see \citet{Yang07} for more details). ``Groups'' can
contain one or more galaxy, and a ``central'' galaxy is simply
identified as the most massive galaxy in each halo (in stars). 

To select quiescent galaxies we use sSFR as a tracer, following
\citet{Franx08}: quiescent galaxies have sSFR $= SFR/
M_{\mathrm{star}}$ smaller than $0.3 \times t^{-1}_{\mathrm{hubble}}
\approx 10^{-11}\ \mathrm{yr}^{-1}$. It is important that we define
quenching by low star formation rate rather than by red colour, given
that one-third of the red galaxies are star forming and the results
are sensitive to this choice as was shown by \citet{Woo13}.

%*****************************************************************************************************
\subsection{The galaxy formation model}\label{model}
%*****************************************************************************************************

\subsubsection{Millennium simulation and cosmology}

To compare observed galaxies with model predictions, we take advantage 
of the publicly available catalogues from the galaxy formation model
presented by \citet{Guo11}. This model was applied to the dark matter
merger trees extracted from the Millennium Simulation
(\citealp{Springel05}). The simulation assumes a WMAP1 cosmology with 
$\Omega_\Lambda = 0.75$, $\Omega_m = 0.25$, $\Omega_b = 0.045$, $n =
1$, $\sigma_8 = 0.9$ and, $h = 0.73$. Note that more recent
measurements of the CMB e.g. with the Planck satellite (Planck
Collaboration et al. 2013) obtain a slightly smaller value for
$\sigma_8 = 0.83$. If only $\sigma_8$ is changed, the present-day
Universe corresponding to a $\sigma_8$ lower than that used in our
simulation can be well approximated by a snapshot corresponding to
some earlier redshift (see e.g. \citealp{Wang08}). The quantitative
results presented in the following (e.g. the fraction of quiescent
central and satellite galaxies) would slightly change. However, we do
not expect the qualitative trends discussed below to be altered
significantly. 

\begin{figure*}
  \centering
  \epsfig{file=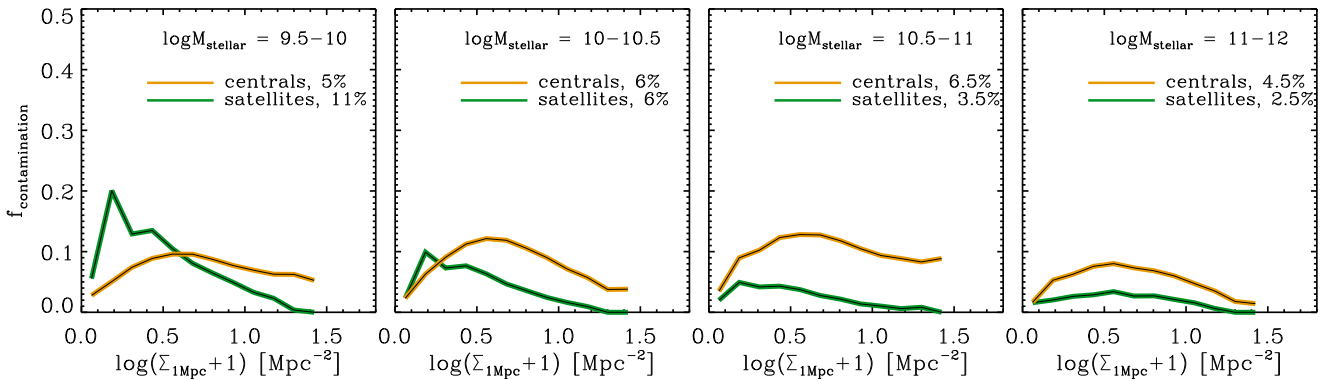, width=1.0\textwidth}
  \caption{The fraction of central (yellow-dashed) and satellite (green-solid)
      galaxies that are misclassified by the algorithm used for the data. These
      are computed by applying the same method used for the observed group
      catalogue (\citealp{Yang07}) to a mock catalogue built from the
      \citet{Guo11} model. Contamination fractions are shown at given stellar
      masses and as a function of the 1~Mpc density. Except for the low-mass
      satellite galaxies, contamination fractions are always below 10 per
      cent.}
 {\label{Conta}}
\end{figure*}

\subsubsection{Environmental processes in the SAM}

The galaxy formation model of \citet{Guo11} includes prescriptions for
gas cooling, re-ionisation, star formation, supernova feedback, metal
evolution, black hole growth, and AGN feedback. This model is is
  based on the earlier models by \citet{DeLucia07} and
\citet{Croton06} but adopts a modified treatment of
  environmental processes acting on satellite galaxies
(strangulation, ram pressure and stellar stripping), and of Supernova
feedback. In addition, the Guo model assumes that galaxies
  effectively become satellites only when they fall within Rvir and
  not - as in the earlier models - as soon as they 
  belong to a larger FOF group implying that environmental effects start
  influencing galaxies at a later time. This definition for a
  satellite galaxy is applied to both type-1 (having a dark matter
  subhalo) and type-2 satellites (orphan galaxies). Furthermore, the
  hot gas mass associated with galaxies being accreted onto larger
  structures is assumed to be reduced by tidal stripping in direct
  proportion to the subhalo's dark matter mass (followed directly in
  the simulations). In addition to tidal forces, the hot gas around
  satellite galaxies is assumed to experience ram-pressure forces due
  to the motion of a satellite galaxy through the inter-cluster
  medium. At the radius where the ram pressure is dominating over the
  self-gravity (to bind the gas), the hot gas is completely
  stripped. The stripping of the hot gas is less efficient than in the
  model of \citet{DeLucia07}, who assumed an instantaneous stripping
  of the hot gas. For more details, we refer the reader to \citet{Guo11}.

\subsubsection{Limitations of the SAM and other models}

The model has been shown to reproduce qualitatively a large variety of
data, both at high redshift and in the local Universe. However, it is not
without problems: even if the model can reproduce the present-day
stellar mass function due to a stronger Supernova feedback than in
former models, it still over-predicts low-mass galaxies at higher
redshifts. In addition, the predicted fraction of red (satellite)
galaxies is still too high with respect to observational measurements. 

These drawbacks are not specific to this particular galaxy formation
model, and rather represent one of the major challenges for recently
published semi-analytic models (e.g. \citealp{Bower06, Somerville08,
  Hirschmann11}; \citealp{Hirsch:12, Bower12, Weinmann12}), as well as for
hydrodynamic simulations (\citealp{Dave11, Weinmann12,
  Hirschmann13a}). There have been some recent attempts to solve these
problems by changing the star formation prescription (\citealp{Wang12}) or by
adopting reincorporation  time-scales which vary inversely with halo
mass (\citealp{Henriques13}),  but none of the proposed solution has
been shown to be completely satisfactory and successful.

\subsubsection{Mimicking observations with the SAM}

To perform a fair comparison between observations and model predictions, we
have projected the cosmological box along the z-axis.  As in the observations,
we use the specific star formation rate as a tracer for the quiescence of a
galaxy. When looking at the distributions of sSFRs at a given stellar
  mass and 1~Mpc density, we find that the adopted cut ($10^{-11}\ yr^{-1}$)
  provides a suitable selection criterion for quiescent galaxies for both
  models and observations.

As a measure for environment, we consider the theoretical halo mass
and the local densities of the model galaxies. For the latter, we
compute the projected density within a cylinder considering two
different radii, $r = 0.2$ Mpc and $r=1$ Mpc using equation
\ref{dens1mpc}.  

To be consistent with the observations, we adopt the same
luminosity cuts in the models: we consider as ``primary'' galaxies
those with a SDSS r-band magnitude down to $M_r < -18$ at $z=0$, and
as ``neighbour'' galaxies only those brighter than $M_r < -20$ within
$\Delta\ v = +/-1000\ \mathrm{km/s}$ of the ``primary'' galaxy (taking
into account the peculiar velocities of the neighbour galaxies). To
assess the environmental histories of our model galaxies we will make
use of the galaxy merger trees, but consider only the main branch of
the trees, i.e. the main progenitors of the $z=0$-galaxies (see Fig.1 in
\citealp{DeLucia12}).

{\subsection{Misclassification of centrals and satellites}

The distinction between central and satellite galaxies in the models is
  based on the dynamical and positional information coming from the simulations:
  by construction, central galaxies are those sitting at the centre of FOF
  groups, and satellites are all other galaxies gravitationally bound to the
  same FOF. In the observations, the central/satellite classification is
  obtained from the group catalogue by \citet{Yang07}. As summarised in
  Section~\ref{observations}, this is built by running a FOF algorithm on the
  data, and then assigning a central/satellite status according to a simple mass
  rank.

Before comparing the observed quiescent central and satellite fractions
  to model predictions, we quantify the contamination fractions due to the
  group classification algorithm used for the data. To this aim, we have
  employed a central/satellite catalogue obtained by applying the
  \citet{Yang07} algorithm for groups detection and  satellite/central
  classification to the galaxies of the Guo model. Fig. \ref{Conta}
  shows the fractions of misclassified centrals (yellow-dashed) and
  satellites (green-solid) at a given stellar mass, as a function of the
  1Mpc-density. We have also specified the total contamination
  fraction in each mass bin (see legend). The contamination of the
  satellite sample is due to misclassified central galaxies as group
  members, which may happen if they reside in the vicinity of a
  group. Instead, contamination of the central sample occurs when the
  group finder incidentally breaks up one group into smaller ones. We
  find no strong dependence of the contamination fractions on density
  (even if for low-mass satellites the contamination fractions tend to
  decline with increasing density). At a given stellar mass bin, the
  fractions are significantly lower than 10 per cent for both centrals
  and satellites (except for low-mass satellites).

Following the approach adopted by \citet{Weinmann09}, we use these
  contamination fractions to estimate the un-contaminated quiescent fractions
  for both centrals and satellites. Briefly, using the contamination
  fractions at a given density and stellar mass $c_{\mathrm{sat/cent}}$, and
  given some observational average quantity $\hat{p}_{\mathrm{sat}}$ and
  $\hat{p}_{\mathrm{cen}}$, one can estimate the corresponding un-contaminated
  average quantities $\bar{p}_{\mathrm{sat}}$ and $\bar{p}_{\mathrm{cen}}$
  from:
\begin{eqnarray}
\hat{p}_{\mathrm{sat}} & = & (1-c_{\mathrm{sat}})
\bar{p}_{\mathrm{sat}} + c_{\mathrm{sat}}\bar{p}_{\mathrm{cen}}\\
\hat{p}_{\mathrm{cen}} & = & c_{\mathrm{cen}}\bar{p}_{\mathrm{sat}} +
(1-c_{\mathrm{cen}}) \bar{p}_{\mathrm{cen}} 
\end{eqnarray}
We find that the quiescent fractions (at a given stellar mass and density)
remain almost unchanged. Thus, in the following analysis, we present results
uncorrected for contamination.

\begin{figure}
  \centering
  \epsfig{file=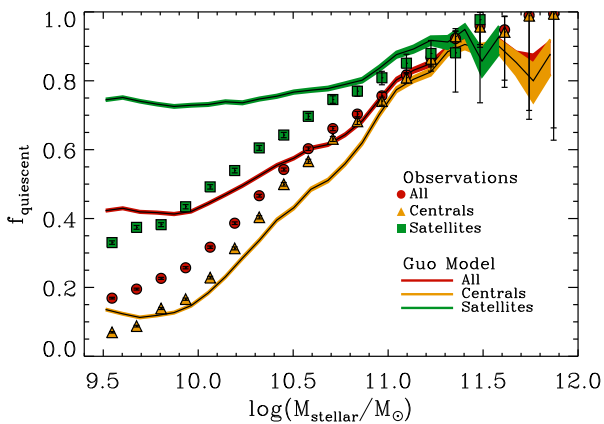, width=0.45\textwidth}
  \caption{Quiescent fraction of all (red-dotted), central (yellow-dashed) and
    satellite (green-solid) galaxies versus the galaxy stellar mass for the
    Guo model (lines and shaded areas) and observations (symbols). At
    fixed stellar mass the quiescent central fraction in the models is
    partly slightly under-predicted, while the amount of quiescent
    satellites is significantly over-predicted. }
 {\label{Quiesc_mass_Guo}}
\end{figure}

%**********************************************************************************
\section{The stellar mass dependence of quiescent galaxies}\label{stellardep}
%**********************************************************************************

The quiescent fraction of galaxies is known to be dependent on both  
stellar mass and density. The dependence on density is strongest for
quiescent low-mass galaxies (see e.g. \citealp{Peng12}). In
Fig. \ref{Quiesc_mass_Guo}, we start by investigating the stellar mass
dependence of the quiescent fraction of all (red-dotted), central
(yellow-dashed) and satellite (green-solid) galaxies in the Guo model
(lines) and in the observations (symbols).  The quiescent fractions
$f_{\mathrm{quiescent}}$ have been calculated as 
\begin{eqnarray}\label{frac}
f_{\mathrm{quiescent}} = n_{\mathrm{qu,x}}/n_{\mathrm{x}},
\end{eqnarray}
where $x$ indicates the (sub-)population of galaxies (all, centrals,
satellites), $n_{\mathrm{qu,x}}$ is the number of quiescent galaxies
and $n_{\mathrm{x}}$ the total number of these galaxies. The
  error bars in these plots are 68\% confidence limits from the Wilson
  (1927) binomial confidence interval (this is the case for all the
  following figures, unless otherwise stated).

\begin{figure*}
  \centering 
  \epsfig{file=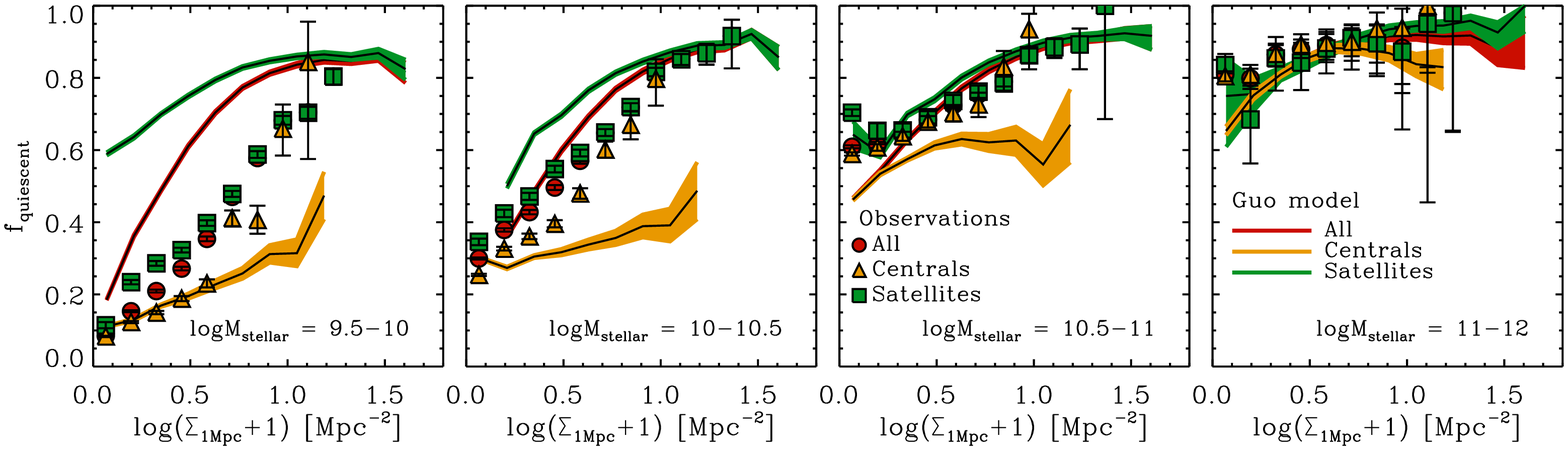,
    width=1.0\textwidth}\vspace{-0.5cm}
  \epsfig{file=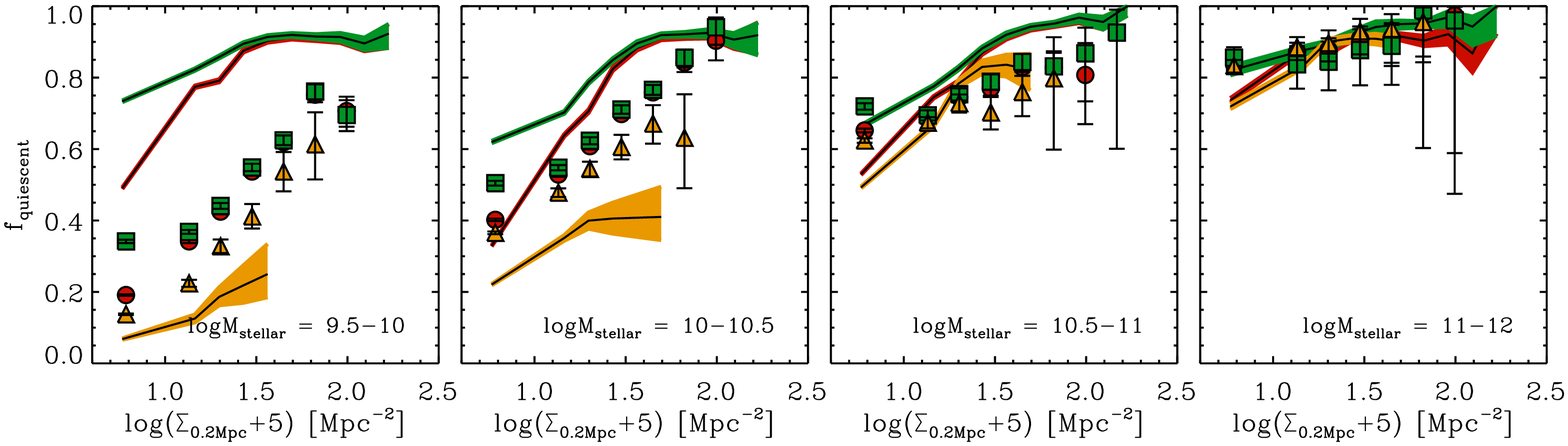,
    width=1.0\textwidth}\vspace{-0.5cm}
  \caption{Quiescent fraction of all (red-dotted), central
    (yellow-dashed) and satellite (green-solid) galaxies versus the
    projected density of a 1~Mpc 
    cylinder (top row) and of a 0.2~Mpc cylinder (bottom row) for
    the observations (symbols) and for the Guo model (lines). Different
    columns correspond to different stellar mass bins as indicated in
    the legend. At stellar masses below $3 \times 10^{10} M_\odot$,
    observations show a strong correlation between the quiescent
    fraction and density of both satellites and centrals. For stellar
    masses above $10^{10}M_{\odot}$ there is 
    only very weak difference between observed centrals and satellites
    which is insignificant contrast to the model predictions.} 
 {\label{Quiesc_dens_Guo}}
\end{figure*}

\begin{figure*}
  \centering 
  \epsfig{file=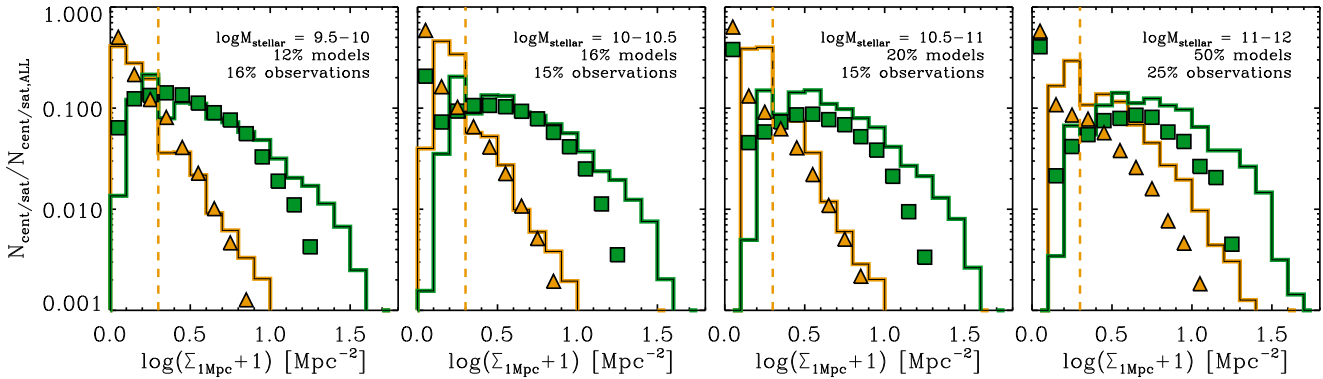,
    width=1.0\textwidth}\vspace{-0.5cm}
  \caption{Distribution of central (yellow-dashed) and satellite (green-solid)
    galaxies (fractions are normalized to the total number of centrals and
    satellites in each mass bin considered) as a function of the 1~Mpc 
    density for different bins of stellar mass. Model predictions are shown 
    as solid lines, while observational data are shown as symbols. The
    agreement between models and observations is very good. The fractions
    given in each panel correspond to the fractions of central galaxies in
    models and observations living at densities above $\log(\Sigma_{1
      \mathrm{Mpc}}+1) =0.3$ (as indicated by the vertical dashed
    line), where data show a strong density dependence for quiescent
    centrals (see Fig.~\ref{Quiesc_dens_Guo}).}   
 {\label{Distr_dens_Guo}}
\end{figure*}

Comparing the observations to the results of \citet{Kimm09} (not shown
in Fig. \ref{Quiesc_mass_Guo}), our observed, quiescent fractions at
galaxy masses below $\log (M_{\mathrm{stellar}} / M_\odot) < 10$ are
somewhat lower than theirs, particularly for satellite
galaxies. \citet{Kimm09} use the Yang catalogue based on the
SDSS data similar to our analysis, adopt the same magnitude cut and
perform a completeness correction. However, they use SFR estimates
based on full SED modelling, while \citet{Brinchmann04} - as used in
our analysis - have estimated the SFR via emission lines. It seems
likely that part of the discrepancy relates to the detection threshold
in UV driving a SFR limit which increases to higher sSFR for lower
mass galaxies. There is a population of galaxies which has
sSFR~$<10^{-11}$ for the SED fitting including UV, but
sSFR~$>10^{-11}$ according to \citet{Brinchmann04}.     

Turning to the model predictions, at galaxy masses below $\log
(M_{\mathrm{stellar}} / M_\odot) < 10.5$, the predictions of the Guo
model significantly over-estimate the observed, quiescent satellite
fraction, while they slightly under-estimate the central one at
intermediate stellar masses. This indicates that the treatment of
central galaxies in the models also needs to be revised. 
In addition, this demonstrates that even the more relaxed, delayed
strangulation assumption in the Guo model (compared to the
instantaneous strangulation) is not sufficient to solve the well-known 
``over-quenching problem'' of satellite galaxies with masses below
$\log (M_{\mathrm{stellar}} / M_\odot) < 10.5$. Overall it shows,
that \textit{the recipes for the physical processes working on both 
  satellite and central galaxies need to be further refined}.

%**********************************************************************************
\section{The dependence of quiescent galaxies on their
  environment}\label{densdep} 
%**********************************************************************************

%**********************************************************************************
\subsection{The fraction of quiescent galaxies as a function of
  density}\label{fqu_dens} 
%**********************************************************************************

In this section, we investigate how model predictions of quiescent
fractions as a function of environment deviate from observations. We
consider the importance of halo mass, central vs satellite status, and
the projected density within a 1~Mpc and a 0.2~Mpc cylinder (as
explained in section \ref{theory}).      

Fig. \ref{Quiesc_dens_Guo} shows the quiescent fraction of all (red-dotted), 
central (yellow-dashed) and satellite (green-solid) galaxies versus
the 1~Mpc density (top row) and versus the 0.2~Mpc density (bottom
row) for the Guo model (lines) and the observations (symbols). We have
divided our central and satellite sample into different stellar mass
bins (as indicated in the legend) in order to separate the dependence
of the quiescent fraction on stellar mass and on  density.   

It is remarkable that in the observations, quiescent fractions at a
given stellar mass and density are very similar for satellite and
central galaxies, and that both fractions reveal a strong dependence
on density for stellar masses below $\log (M_{\mathrm{stellar}} /
M_\odot) < 10.5$. 

\begin{figure*}
  \centering
  \epsfig{file=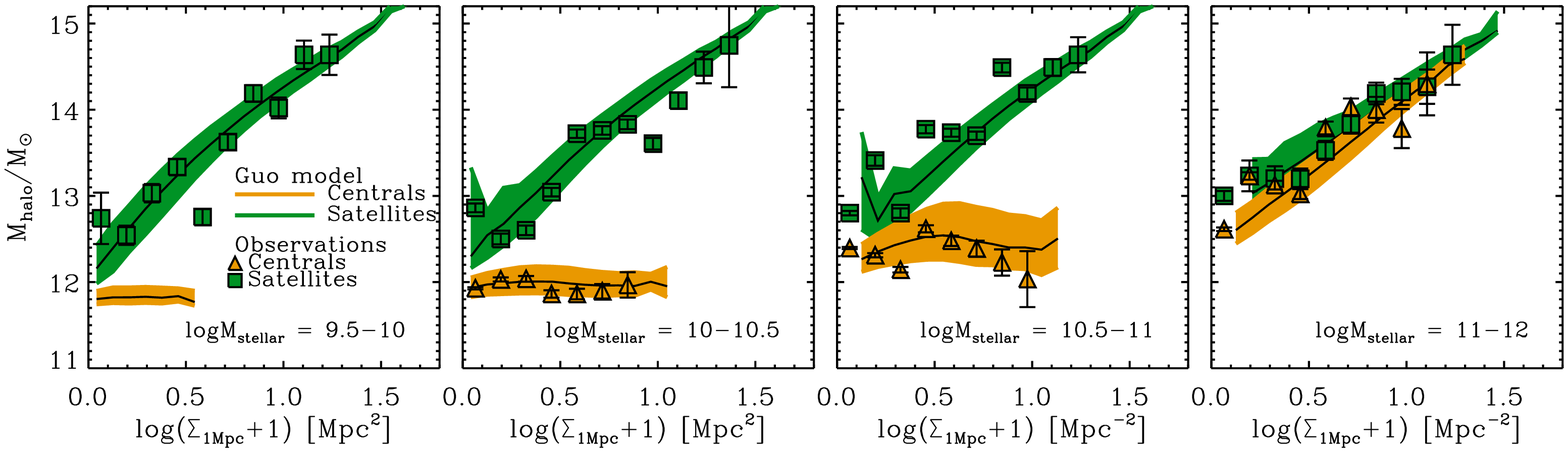, width=1.0\textwidth}\vspace{-0.5cm} 
  \epsfig{file=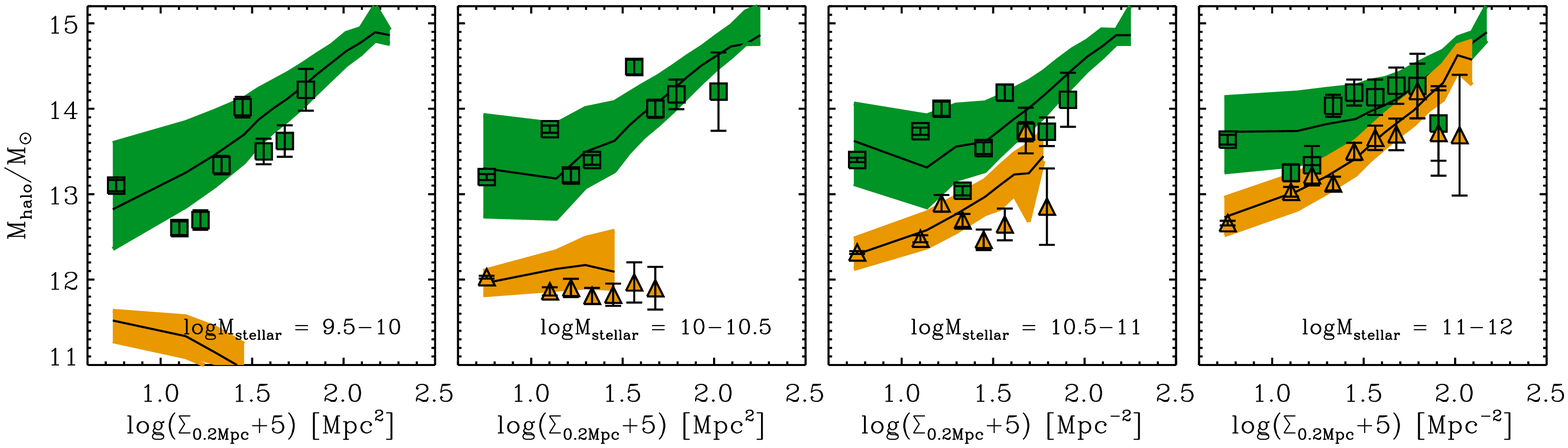, width=1.0\textwidth}\vspace{-0.5cm} 
  \caption{Median parent halo mass versus the 1~Mpc density (top row)
    and versus the 0.2~Mpc density (bottom row) in different stellar
    mass bins (different columns) for the Guo model (shaded areas)
    and for the observations (symbols).  We also distinguish between
    satellites (green areas, solid lines and squares) and centrals
    (yellow areas, dashed lines  and triangles). The shaded areas and
    the error bars of the symbols refer to the 25th and 75th
    percentiles, respectively. Observations and models are in good
    agreement: the parent halo mass of satellites scales with their
    density, but interestingly, the halo mass of centrals correlates
    with density \textit{only} for the most massive galaxies (right
    column). }  
 {\label{Halocent_dens}}
\end{figure*}

In contrast to the observations, the modelled quiescent fractions 
reveal significantly larger differences between
satellites and centrals.  For satellites, the agreement between
observations and model predictions is reasonably good for massive
galaxies ($\log (M_{\mathrm{stellar}} / M_\odot) > 10.5$) and for
those in high-density regions ($\log(\Sigma_{1\mathrm{Mpc}}+1)>1
[\mathrm{Mpc}^{-2}]$).  Apart from that, at a given density and
stellar mass below $\log (M_{\mathrm{stellar}} / M_\odot) < 10.5$,
the Guo model significantly over-predicts the fraction of quiescent 
satellites. The disagreement gets worse with decreasing galaxy
mass and decreasing density. In addition, for
satellites with stellar masses below $\log (M_{\mathrm{stellar}} /
M_\odot) < 10.5$ the model predicts a weaker dependence on both the
0.2~Mpc and the 1~Mpc density than what is observed. 

For centrals, the dependence of the quiescent fraction on the 1~Mpc
density is weaker in the models than in the observations, while the
dependence on the 0.2~Mpc density has similar strength. The model also 
predicts realistic quiescent central fractions at high stellar masses
$\log (M_{\mathrm{stellar}} / M_\odot) > 11$ or at very low densities.
Otherwise, at higher densities and for lower stellar masses, the Guo
model under-estimates the amount of quiescent centrals: the mismatch
becomes stronger for centrals with decreasing galaxy masses and
increasing densities.

We find that the offset between model predictions and observations at fixed 
density is driven by the contribution of haloes with mass 
$\log (M_{\mathrm{halo}} /   M_\odot) < 11.7$, where the fraction of quiescent 
model centrals is very low (it rapidly increases for more massive haloes).
Since these low mass haloes are close to the resolution limit of the  
Millennium simulation, we have tested and confirmed that model quiescent 
fractions are not affected by resolution using the Guo model applied to the
Millennium-II simulation (this corresponds to a smaller cosmological volume,
but a higher resolution). The disagreement found indicates that also the
treatment of model central galaxies needs to be improved (this is in agreement 
with previous studies, see e.g. \citealt{Cucciati12}).

Comparing the 1~Mpc density with the smaller scale 0.2~Mpc density in 
observations, for both the central and satellite quiescent fractions
with stellar masses below  $\log (M_{\mathrm{stellar}} / M_\odot) < 11$, the
dependence on the 0.2~Mpc density is slightly weaker than the one on the
1~Mpc density (the relative increment of the quiescent satellite
fractions from low density to high density, $=f_{\mathrm{qu,sat}}
(\mathrm{high}\ \Sigma) / f_{\mathrm{qu,sat}}(\mathrm{low}\ \Sigma)$
is larger for the 1~Mpc than for the 0.2~Mpc density ). Also model
quiescent satellites show a slightly weaker dependence on the 0.2~Mpc
density than on the 1~Mpc one (this will be discussed further in section
\ref{centrals}).    

To understand how important the environmental dependence is in the 
population of central galaxies, Fig. \ref{Distr_dens_Guo} shows the 
distributions of central (yellow) and satellite (green) galaxies (normalised 
to the total number of centrals and satellites in each mass bin) as a function
of the 1~Mpc density for different stellar mass bins (different panels). Data
are shown as symbols, while model predictions are shown by solid lines. The
agreement between model predictions and observations is generally very
good, particularly for the lowest stellar mass bin considered, even if for
the more massive galaxies, models overpredict the number of satellites
at $\log(\Sigma_{1\mathrm{Mpc}}+1) > 3.0$ and they overpredict the
fraction of massive centrals at all densities by a factor ~2. The
fractions of  central galaxies residing at densities
$\log(\Sigma_{1\mathrm{Mpc}}+1) > 0.3$  vary between 15 to 25 per cent
in the data (the largest fractions are found  in the highest stellar
mass bin) and between 12 to 50 per cent in the models. The bulk of
central galaxies is residing at low densities, while satellite
galaxies dominate the high-density regions. Therefore, the density
dependence  of the overall galaxy population is driven by satellite
galaxies. There is,  however, a non negligible fraction of central
galaxies residing in high-density regions, and their quiescent
fractions increases with increasing density.

%**********************************************************************************
\subsection{Relation between density and halo mass}\label{density_halo}
%**********************************************************************************

In this section, we explore the \textit{theoretical} interpretation of
the observed density dependence (particularly on the 1~Mpc density) of
both centrals and satellites. We start by investigating whether the
density dependence reflects a dependence on halo mass, i.e. we analyse
whether and how strongly the 1~Mpc and the  0.2~Mpc density is related
to the halo mass of centrals and the parent halo mass of satellites.

Fig. \ref{Halocent_dens} shows the median halo mass versus the
1~Mpc density (top row) and versus the 0.2~Mpc density (bottom row)
for centrals (yellow-dashed) and satellites (green-solid) in models
(lines) and observations (symbols). The shaded areas indicate the
range between the 25th and 75th percentile of the distribution of halo
masses at fixed density. For both centrals and satellites, we find a
good agreement between models and observations. This analysis extends
a recent study of \citet{Haas12}, who also investigate the relation
between density (using different estimators) and halo mass (see also
\citet{Muldrew12}), but \textit{they do not distinguish between
  centrals and satellites or different stellar mass bins}. They find
that the density (for fixed aperture densities) is \textit{always}
correlated with halo mass.   

Our results are consistent with theirs when we consider satellite
galaxies: the parent halo mass of satellites is strongly correlated
with density irrespective of the density scale and the stellar
mass. The median parent halo masses are above $\log (M_{\mathrm{halo}}
/ M_\odot) = 12.5$ corresponding to virial radii larger than $>
0.2$~Mpc.  

The correlation between parent halo mass and density is slightly
weaker for the 0.2~Mpc scale than for the  1~Mpc scale (the median
parent halo masses at low 0.2~Mpc densities are higher than at low
1~Mpc densities). This is likely due to the fact that the small-scale
0.2~Mpc density preferentially traces ``intra-halo'' scales
(i.e. scales smaller than the virial radius) of parent haloes 
more massive than $\log (M_{\mathrm{halo}} / M_\odot) > 12.5$. 
 In contrast, the 1~Mpc density will capture the entire haloes up to
masses of $\log (M_{\mathrm{halo}} / M_\odot) = 15$ corresponding to
virial radii of roughly 1~Mpc.

This behaviour explains some of the trends shown in
Fig. \ref{Quiesc_dens_Guo}, where the dependence of the quiescent
satellite fraction on the 0.2~Mpc density was shown to be slightly 
weaker than the one on the 1~Mpc density. This indicates that
\textit{densities of intra-halo scales do not  capture the total
  effect of environment on quiescent satellites}.

In contrast to the satellites, the halo mass of central galaxies
(yellow areas in Fig. \ref{Halocent_dens}) \textit{depends on the
  1~Mpc density (resp. the 0.2~Mpc density) only for massive galaxies}
with $\log (M_{\mathrm{stellar}} / M_\odot) > 11$ (resp. $\log 
(M_{\mathrm{stellar}} / M_\odot) > 10.5$). In other words, a correlation 
between the 1~Mpc density (resp. 0.2~Mpc density) and halo mass only
emerges for galaxies with a median halo mass above $\log
(M_{\mathrm{halo}} / M_\odot) \sim 13$ (resp. $\log (M_{\mathrm{halo}}
/ M_\odot) \sim 12.5$). Such halo masses correspond roughly to virial
radii larger than $\sim 300$~kpc (resp. $\sim 200$~kpc). This means
that the 1~Mpc density probes \textit{super-halo scales} for haloes
with $\log (M_{\mathrm{halo}} / M_\odot) \sim 13$. For smaller haloes
(in which low mass centrals are preferentially sitting, $\log
(M_{\mathrm{halo}} / M_\odot) \sim 11-12$ with $r_{\mathrm{vir}} \leq
150$~kpc), 1~Mpc and 0.2~Mpc densities do not correlate with the  
halo masses, neither in models nor in observations. This reflects the
averaging over scales larger than the corresponding virial radii.

To summarise this section, the strong correlation between density and
host halo mass for satellites implies that the dependence of the
quiescent satellite fractions on the 1~Mpc density (see top row of
Fig. \ref{Quiesc_dens_Guo}) can be theoretically understood - at least
partly - in terms of their dependence on parent halo mass. However,
the correlation between density and halo mass does not allow for any
conclusion about the fundamental relation between environment and the 
quiescent fraction. We will discuss this in more detail in section
\ref{satellites}. In contrast, given that density and halo mass are
unrelated for centrals (Fig. \ref{Halocent_dens}), \textit{the density
  dependence of quiescent central fractions, particularly those of
  lower mass galaxies, can hardly originate from a dependence on halo
  mass.}    

\begin{figure*}
  \centering
  \epsfig{file=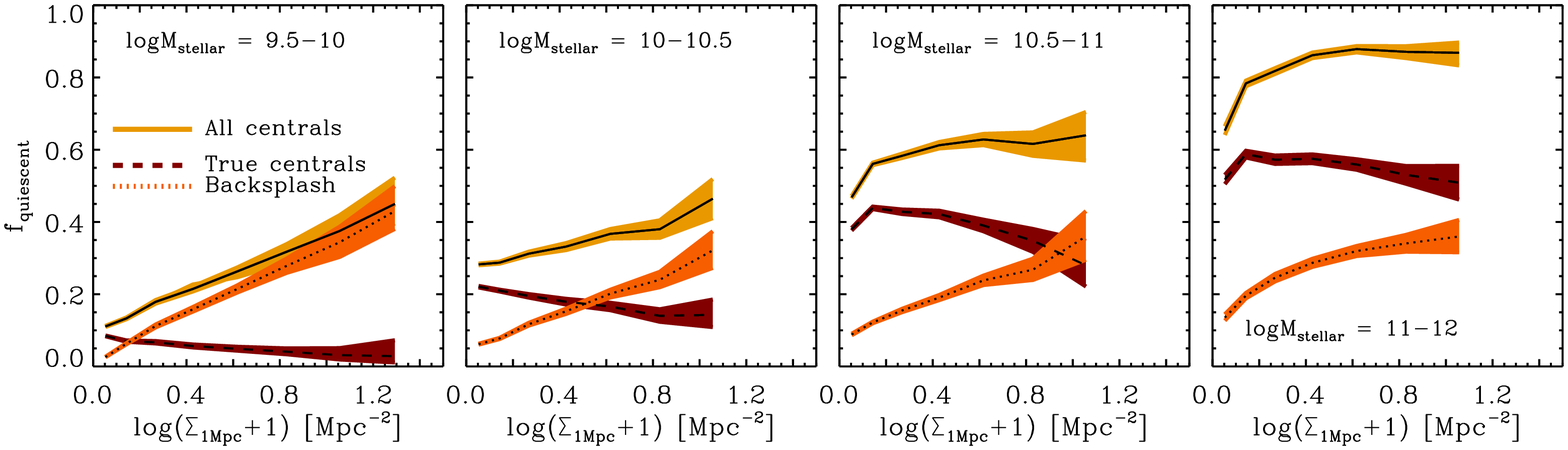, width=1.0\textwidth}\vspace{-0.5cm} 
  \epsfig{file=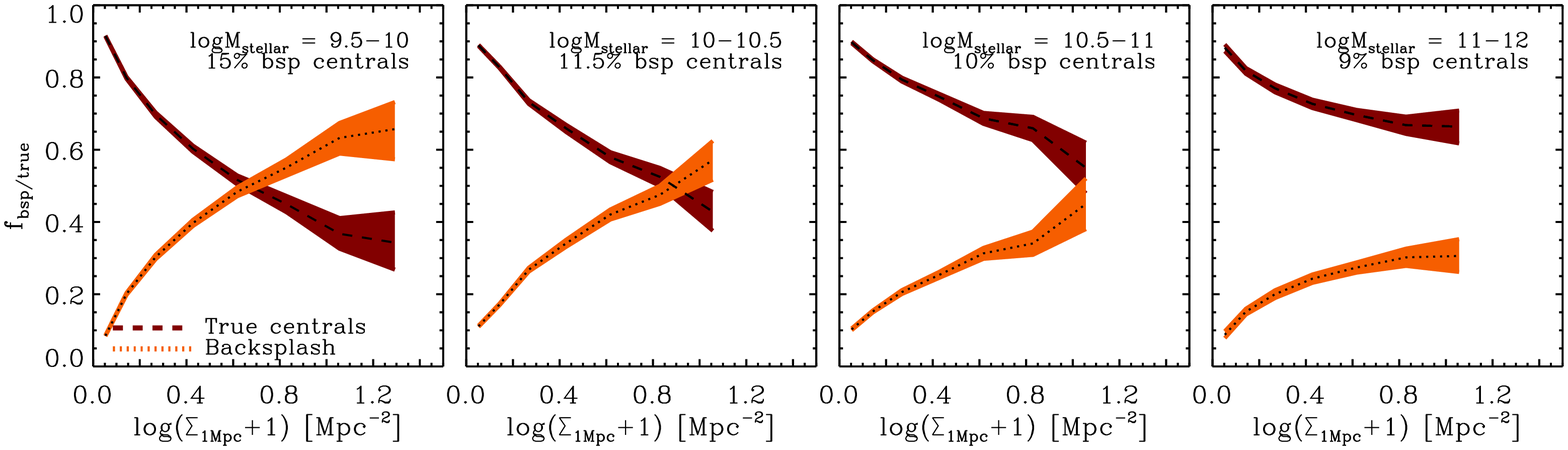, width=1.0\textwidth}\vspace{-0.5cm} 
  \caption{Top row: the quiescent fraction of model central galaxies
    (solid lines with yellow areas) in different stellar mass bins
    (different columns) versus the 1~Mpc density distinguishing
    between centrals having always been centrals (true centrals,
    dashed lines with dark red areas) and centrals having been
    satellites in the past (backsplash population, dotted lines with
    orange shaded areas). The quiescent fractions of backsplash
    centrals are responsible for the overall density dependence of
    quiescent model centrals. Bottom row: The total fractions of
    backsplash (orange) and true (dark red) central galaxies versus
    the 1~Mpc density in different stellar mass bins. The lower the galaxy
    stellar mass and the higher the density, the higher is the probability
    that a central galaxy was a satellite galaxy in the past.}  
 {\label{Quiesc_cent_bp}}
\end{figure*}

%**********************************************************************************
\subsection{Central galaxies}\label{centrals}
%**********************************************************************************

We have shown in section \ref{density_halo} that the halo mass of
centrals can hardly be responsible for the density dependence of
low-mass quiescent centrals. This result may indicate the existence of 
environmental effects on super-halo scales affecting central
galaxies. There are two possible explanations for this. 

First, present-day central galaxies may have been satellites in the
past and therefore, they can have experienced environmental processes
like strangulation for some fraction of their life time. Such satellites have
probably been on highly eccentric orbits so that - after one or more
pericenter passages - they could have left their parent halo and
become central galaxies again (outside the virial radius of the 
previous parent halo). We refer to these present-day central galaxies
as ``backsplash'' centrals. Strictly speaking, of course, such
backsplash centrals are not affected by environment of super-halo
scales but have been subject to environmental effects while orbiting
within a more massive halo in the past.    

The galaxy formation model accounts for such a backsplash effect
automatically as it uses the kinematic input from dark matter
simulations (i.e. it follows the orbits from the simulations): as long
as the backsplash central galaxy is a satellite, it is assumed to be
affected by tidal and ram pressure stripping. After ejection from the
parent halo, the backsplash central is treated as a ``normal'' central
galaxy which means that such a galaxy does not experience any
environmental effects anymore. Instead, it can (re-)accrete gas which
can cool and form stars. Nevertheless, its hot halo content is reduced
due to the time spent as satellite and thus, a backsplash central
galaxy evolves differently to if it had always been a central galaxy
in the past.  

Alternatively, central galaxies, which have never been satellites in
the past, might suffer environmental effects on super-halo scales.
This was nicely demonstrated in recent study of \citet{Bahe13} using
hydrodynamical simulations. They found that a direct interaction with
an extended hot gas 'halo' of a group or cluster can be sufficiently
strong to strip the hot gas atmospheres of infalling galaxies as far
out as $\sim 5 \times r_{\mathrm{vir}}$. However, the hot gas
stripping was not found to significantly affect the on-going star
formation and the quiescent fraction of galaxies outside the virial
radius. It may, thus, be expected to have only a minor impact. So far,
there is no recipe in the models to account for this second effect.

%**********************************************************************************
\subsubsection{Backsplash population}
%**********************************************************************************

To quantify the statistical relevance of the backsplash population in
the models (note that this information is of course not accessible in
the observations), we make use of the galaxy merger trees of the Guo
model and trace the main progenitors of the central galaxies back in
time to analyse whether they have been a satellite in the past. 

The top row of Fig. \ref{Quiesc_cent_bp} shows the 1~Mpc density
dependence of the quiescent fraction of all model centrals (solid
lines with yellow shaded areas as shown before in Fig.
\ref{Quiesc_dens_Guo}). We additionally distinguish between quiescent
centrals having \textit{always} been centrals, i.e. \textit{``true''}
centrals, (dashed lines with dark red shaded areas) and quiescent
centrals having been \textit{satellites in the past},
i.e. \textit{``backsplash''} centrals (dotted lines and orange shaded
areas). For the true and backsplash centrals, we have calculated their
quiescent fractions with respect to the \textit{total amount of
  centrals}, i.e.:  
\begin{eqnarray}\label{bspfrac}
f_{\mathrm{quiescent}} = \frac{n_{\mathrm{qu,bsp/true}}}{n_{\mathrm{cent}}},
\end{eqnarray}
where $n_{\mathrm{qu,bsp/true}}$ is the amount of quiescent backsplash
or true centrals and $n_{\mathrm{cent}}$ the total number of
centrals. 

This fraction is a super-position of the quiescent fraction with
respect to the amount of only backsplash/true centrals
($=n_{\mathrm{qu,bsp/true}}/n_{\mathrm{bsp/true}}$) and the fraction
of backsplash/true centrals ($= n_{\mathrm{bsp/true}} /
n_{\mathrm{cent}}$):  
\begin{equation}\label{suppos}
f_{\mathrm{quiescent}} = \frac{n_{\mathrm{qu,bsp/true}}}{n_{\mathrm{cent}}} =
\frac{n_{\mathrm{qu,bsp/true}}}{n_{\mathrm{bsp/true}}} \times \frac{n_{\mathrm{bsp/true}}}{
n_{\mathrm{cent}}},
\end{equation}
where $n_{\mathrm{bsp/true}}$ is the total amount of true/backsplash
central galaxies. 

The top row of Fig. \ref{Quiesc_cent_bp} shows that the quiescent
fraction of true central galaxies is not or even negatively correlated
with the 1~Mpc density, while the one of the backsplash population is
strongly dependent on density. The bottom row of Fig.
\ref{Quiesc_cent_bp} shows the fraction of backsplash and true
centrals ($= n_{\mathrm{bsp/true}} / n_{\mathrm{cent}}$) versus  the 
1~Mpc density in different stellar mass bins. It is clear from the
figure that the density dependence of the quiescent fraction of
backsplash centrals is mainly caused by a strong positive density
dependence of the fraction of backsplash centrals. 

In the bottom row of Fig.~\ref{Quiesc_cent_bp} we also give the 
fractions of backsplash galaxies (with respect to the total number of 
central galaxies) in each stellar mass bin. About 15 per cent of the 
lowest mass centrals have been satellites in the past. At fixed stellar
mass, the fraction of backsplash centrals increases with increasing
density so that e.g. for low-mass centrals residing at densities 
$\log (\Sigma_{1\mathrm{Mpc}} + 1) >0.3$ at least 30 per cent of centrals 
have been satellites in the past. For larger stellar masses, the fraction 
of backsplash galaxies decreases to about 9 per cent. For galaxies with
stellar mass $\log (M_{\mathrm{halo}} / M_\odot) > 11$, the majority of the
central galaxies have been centrals for their entire life (almost 
independently of density).

Our fractions of backsplash centrals are somewhat larger than those
found in a recent study by \citet{Wetzel14}: their fractions of 
ejected/backsplash galaxies is below 10 per cent while we find fractions up 
to about 15 per cent. We believe that these differences can be ascribed to 
differences in the algorithm used to identify dark matter haloes and to
construct their merger trees (and probably also in the definition for
satellite galaxies in the Guo model). Although quantitatively
different, however,  the trends found for our model galaxies are
qualitatively consistent with those found by \citet{Wetzel14}.

\begin{figure}
  \centering
  \epsfig{file=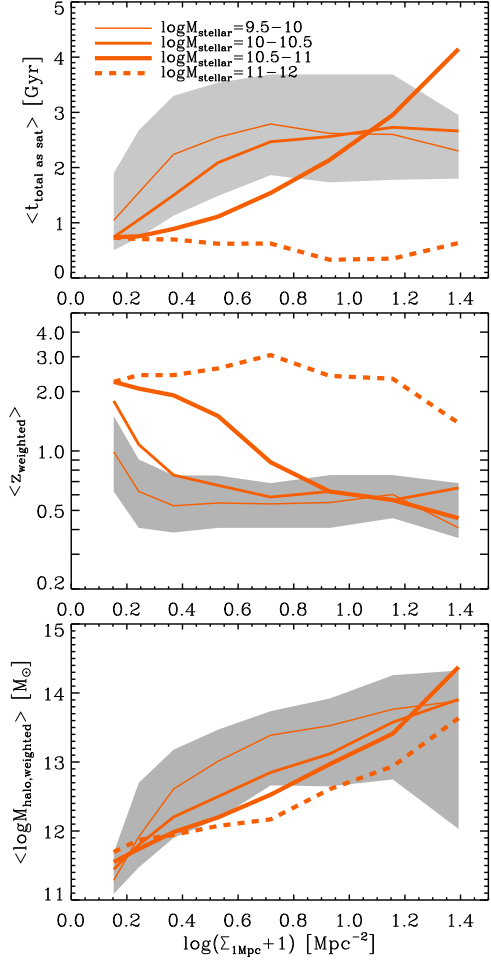,
    width=0.4\textwidth} 
  \caption{Top panel: average total time backsplash centrals have spent
    as a satellite versus the present-day 1~Mpc density for
    different stellar mass bins (indicated by different line
    styles). Middle panel: median weighted redshift, at which
    backsplash centrals have been satellites in the past, versus
    the present-day density for different stellar mass bins. Bottom
    panel: median weighted parent halo mass, in which backsplash
    centrals have been residing as satellites, versus present-day
    density for different stellar mass bins. The grey shaded areas
    always illustrate the 25 and 75 percentiles of the different
    quantities for the lowest stellar mass bin.}     
 {\label{Mean_bp}}
\end{figure}

In Fig. \ref{Mean_bp}, we analyse the density and stellar mass
dependence of some properties of backsplash central galaxies in the
models: the top panel shows the the median total time
$t_{\mathrm{total}}$ which backsplash centrals have spent as
satellites,  the middle panel illustrates the weighted median or
effective redshift $z_{\mathrm{weighted}}$ at which backsplash
centrals have been satellites in the past, and the bottom panel shows
the weighted median or  effective parent halo mass in which backsplash
centrals have been residing as satellites - always distinguishing
between different stellar mass bins (different line styles). The grey
shaded areas illustrate the 25 and 75 percentiles of the
different quantities for the lowest stellar mass bin. The dispersion
is similar for the higher stellar mass bins.

The weighted or effective redshift for a backsplash central galaxy
is estimated as:  
\begin{eqnarray}\label{z_weighted}
z_{\mathrm{weighted}} = \frac{\sum_{\substack{i}}\left(  \bar{z}_i
    \times \Delta t_i \right)}{t_{\mathrm{total}}},
\end{eqnarray}
where $t_{\mathrm{total}}$ is the total amount of time the galaxy has
spent as a satellite, $\Delta t_i$ is the time interval during which
the galaxy has continuously been a satellite (without any
interruption of being a central) and $\bar{z}_i$ is the corresponding
average redshift. The weighted halo masses for a backsplash central
have been calculated in the same manner (replacing $z_i$ with
$M_{\mathrm{halo,}i}$). Such weighted quantities are supposed to give
an estimate for the ``typical'' redshift, at which backsplash centrals
have most likely been satellites, or the mass of a ``typical'' parent
halo, in which backsplash centrals have been residing in the past. 

The top panel of Fig. \ref{Mean_bp} shows that, on average, backsplash
centrals have been satellites for 1-3~Gyr depending on their stellar
mass and environment. Generally, the higher the present-day density
is, higher is the average total time backsplash centrals have spent as
satellites. This is a consequence of hierarchical clustering:
present-day high-density regions have most likely emerged out of
over-dense regions in the past and higher densities give a higher
probability of being temporarily accreted onto a massive halo. 

For a given density below $\log(\Sigma_{1\mathrm{Mpc}}+1) < 1\
[\mathrm{Mpc}^{-2}]$, low mass backsplash centrals have been
satellites for a somewhat longer time than more massive centrals. This  
may be a consequence of the orbits: low mass backsplash galaxies on a
pure radial orbit move to a larger apo-centre (and thus to a region of
lower density) than higher mass ones because dynamical friction is
weaker for them. Any high mass backsplash galaxies which make it out
to these regions of low density despite the strong effects of dynamical
friction are either escaping a low mass host halo, or/and just passing
by with a relatively large impact parameter. At high density none of
this applies, and therefore, more massive backsplash
centrals/satellites are likely to have been in more massive haloes for
a longer time.  

The middle panel of Fig. \ref{Mean_bp} shows that backsplash centrals
have preferentially been satellites at redshifts between
$z_{\mathrm{weighted}} \sim 0.5-2$. Backsplash centrals of similar
stellar mass in high-density regions have been satellites at later
times than those in low-density regions. Backsplash centrals in
high-density regions have been on average satellites only 5~Gyr ago
($z\sim0.5$), while those at low-density regions have been satellites
9~Gyr ago ($z \sim 2$).   

At a given density below $\log(\Sigma_{1\mathrm{Mpc}} +1) < 1\
[\mathrm{Mpc}^{-2}]$, more massive backsplash centrals have on average
been satellites at higher redshifts, i.e. a longer time ago, while
less massive ones have preferentially been satellites at later times. 
As a consequence of hierarchical clustering (more massive haloes form
later than low mass ones), low-mass backsplash centrals have been residing in
more massive parent haloes than those with high stellar masses and/or
those residing at low densities (see bottom panel of Fig. \ref{Mean_bp}).

To summarise, these results indicate that environmental effects on
central galaxies due to the 'backsplash' effect are particularly
important for low-mass centrals and for centrals at high densities, as
they have typically been satellites more recently. Instead,
present-day massive backsplash centrals at low densities may have
likely been less affected by environmental processes because they
have been satellites in low mass haloes, when the Universe had only
one third of its current age. Since then, they have been evolving like
a ``normal'' central galaxy. 

\begin{figure}
  \centering
  \epsfig{file=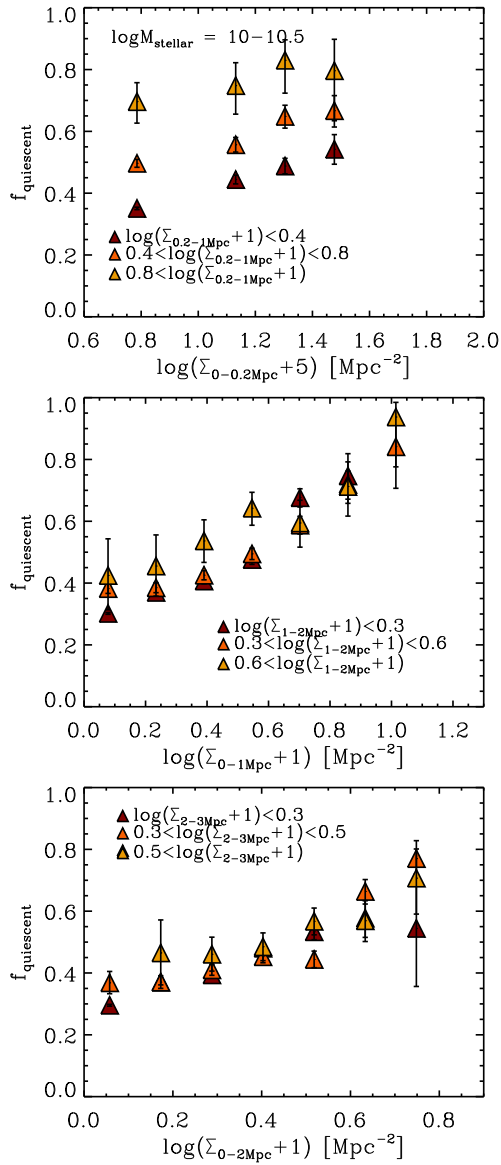,
    width=0.4\textwidth}\vspace{-0.3cm} 
  \caption{The quiescent fractions of observed central galaxies with 
    masses of $\log M_{\mathrm{gal}} = 10-10.5$ versus the
    smaller-scale 0-0.2~Mpc (upper panel), 0-1~Mpc (middle panel) and
    0-2~Mpc density (lower panel) binned in different larger-scale,
    non-overlapping annulus densities (0.2-1~Mpc, 1-2~Mpc and 2-3~Mpc, 
    respectively, indicated by different colours and increasing line
    thickness with decreasing density). Observations mainly
    reveal a residual effect on the quiescent centrals on scales
    between 0.2-1~Mpc but hardly beyond 1~Mpc.}   
 {\label{Quiesc_cent_multi-scale}}
\end{figure}

%**********************************************************************************
\subsubsection{The spatial extent of environmental effects on super-halo scales}
%**********************************************************************************

In the previous section, we  have demonstrated that for model galaxies 
the increasing fraction of quiescent centrals as a function of density
is partially driven by an increasing fraction of the backsplash
galaxies.  Observations, however, reveal a somewhat stronger
density dependence of quiescent centrals than the models (see 
Fig. \ref{Quiesc_dens_Guo}). This likely indicates the need for a
strengthened influence of environment on central galaxies or a refined 
treatment of internal processes.  

Unfortunately, observations do not allow the significance of a
backsplash population and their contribution to the overall
environmental dependence of centrals to be probed directly.
Observations can, however, be used to estimate out to which scales
environmental effects work on centrals and they can help to assess the
scales on which  environment is influencing central galaxies most
efficiently.  

For such an analysis, we follow the multi-scale approach presented
in a paper by \citet{Wilman10}. They calculate the density of two
formally independent, non-overlapping annuli: the one of a
smaller-scale cylinder and the one of a larger-scale annulus. This 
allows to extract and to analyse residual effects of the large-scale
density on quiescent galaxies on top of the small-scale
density. The annulus density is computed as:     
\begin{equation}
\Sigma_{ri-ra\mathrm{Mpc}} = \frac{N(ra)- N(ri)}{\pi (ra^2-ri^2)},
\end{equation}
where $N(ra)$ is the amount of neighbours within a cylinder of radius
$ra$ and $N(ri)$ the one within a cylinder of radius $ri$. 

\begin{figure*}
  \centering
  \epsfig{file=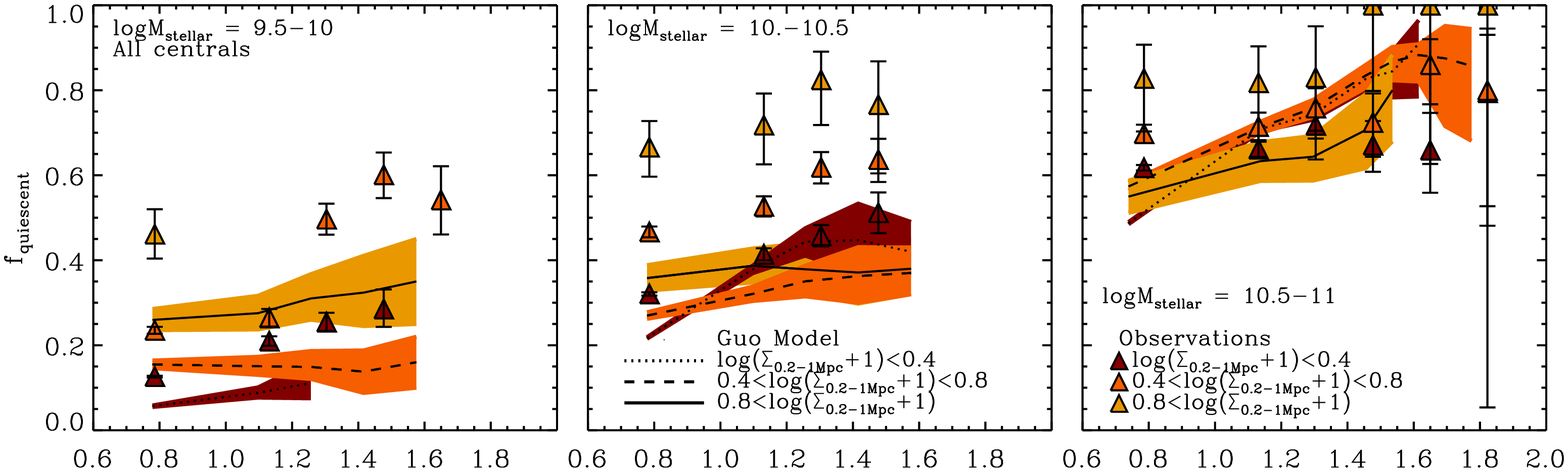,
    width=1.0\textwidth}\vspace{-0.8cm} 
  \epsfig{file=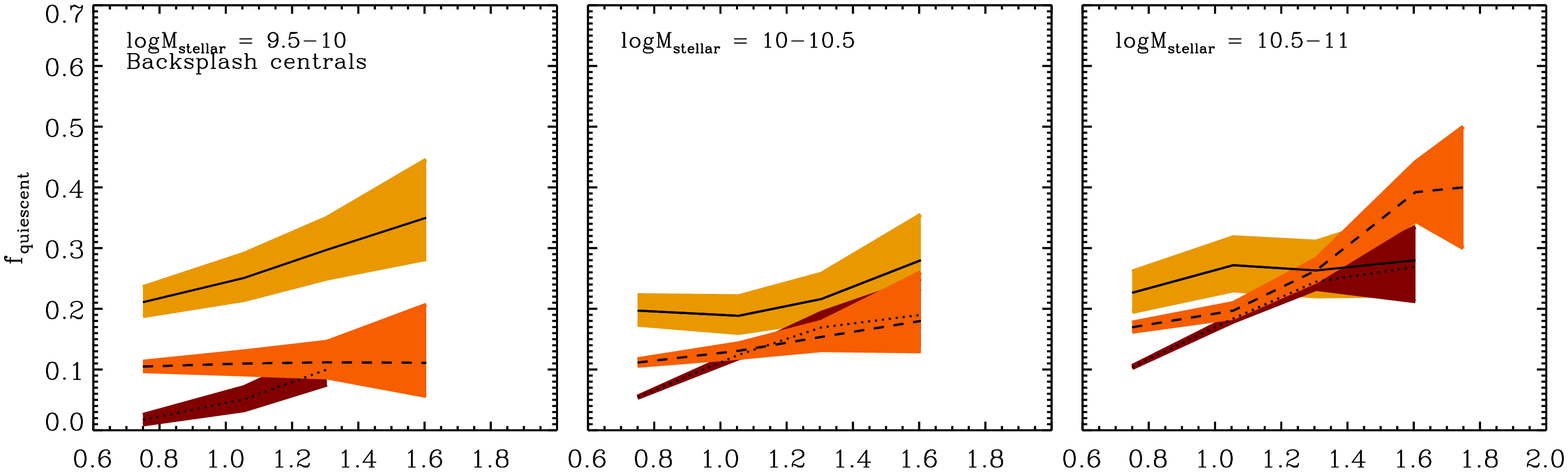,
    width=1.0\textwidth}\vspace{-0.8cm} 
  \epsfig{file=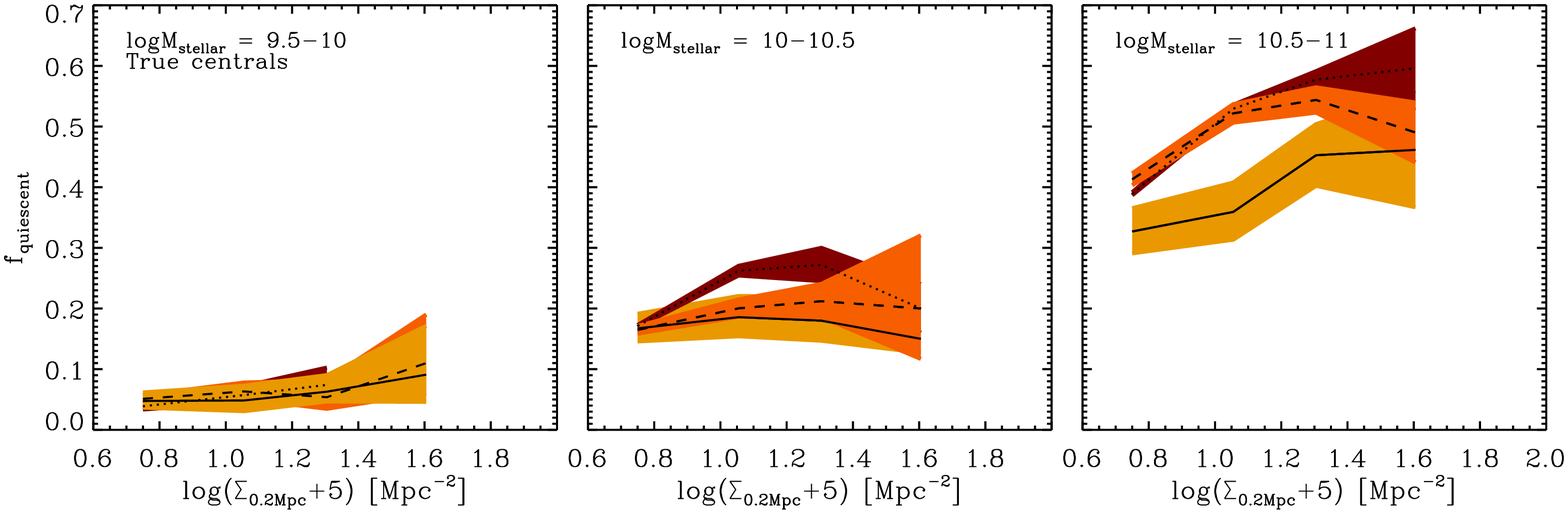,
    width=1.0\textwidth}\vspace{-0.3cm}  
  \caption{Top row: quiescent fractions of central galaxies are
    plotted versus the small-scale 0-0.2~Mpc density for different
    large-scale 0.2-1~Mpc density bins (different colours) in
    observations (symbols, increasing line thickness with decreasing
    density) and models (lines with shaded areas). Middle and bottom
    rows: Same as the top row, but now dividing the quiescent central
    galaxies into backsplash and true centrals, respectively. The
    quiescent fractions are calculated  with respect to the total
    amount of central galaxies. In the models, any residual effect of
    density on super-halo scales is caused by the backsplash
    population of central galaxies.}   
 {\label{Quiesc_cent_sh}}
\end{figure*}

Fig. \ref{Quiesc_cent_multi-scale} shows the quiescent central
fraction as a function of the smaller-scale cylindrical density binned
in larger-scale and non-overlapping annulus densities (different
colours). We have considered three different cylinder-annulus pairs of
0-0.2~Mpc \& 0.2-1~Mpc (top panel), 0-1~Mpc \& 1-2~Mpc (middle panel)
and 0-2~Mpc \& 2-3~Mpc (bottom panel). This is shown only for
$\log (M_{\mathrm{stellar}}/ M_\odot) = 10 - 10.5$ galaxies, but
the qualitative trends do not change when varying the stellar mass
bin. Observations clearly indicate a strong residual effect of the
0.2-1~Mpc scales on top of the 0.2~Mpc density (top panel): the higher
the (larger-scale) annulus density at a given (smaller-scale)
cylindrical density is, the higher is the quiescent
fraction. Observations do not reveal any significant residual trend
for scales beyond 1~Mpc. This is in agreement with previous
observational results of \citet{Wilman10}.

\begin{figure*}
  \centering
  \epsfig{file=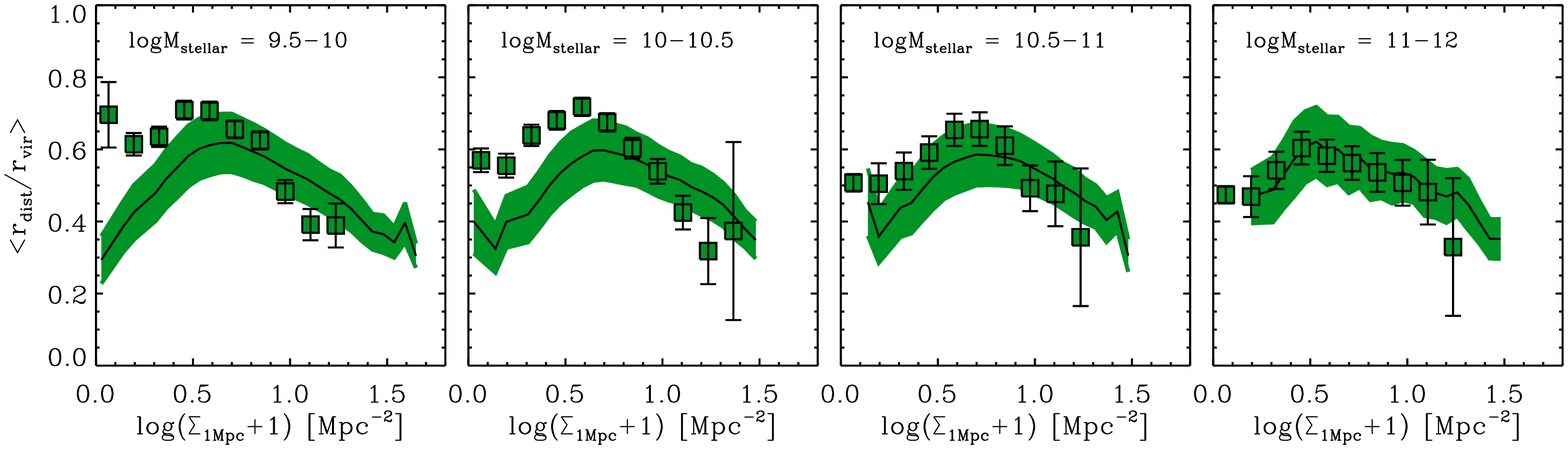,
    width=1.0\textwidth}\vspace{-0.5cm}  
  \caption{Mean radial distance versus the 1~Mpc density for
    satellites in different stellar mass bins (different panels) in
    observations (symbols) and models (lines with shaded
    areas). Density and radial 
    distance are strongly correlated for densities above $\log
    (\Sigma_{1\mathrm{Mpc}} +1) > 0.5$.}  
 {\label{Dens_distsat}}
  \centering
  \epsfig{file=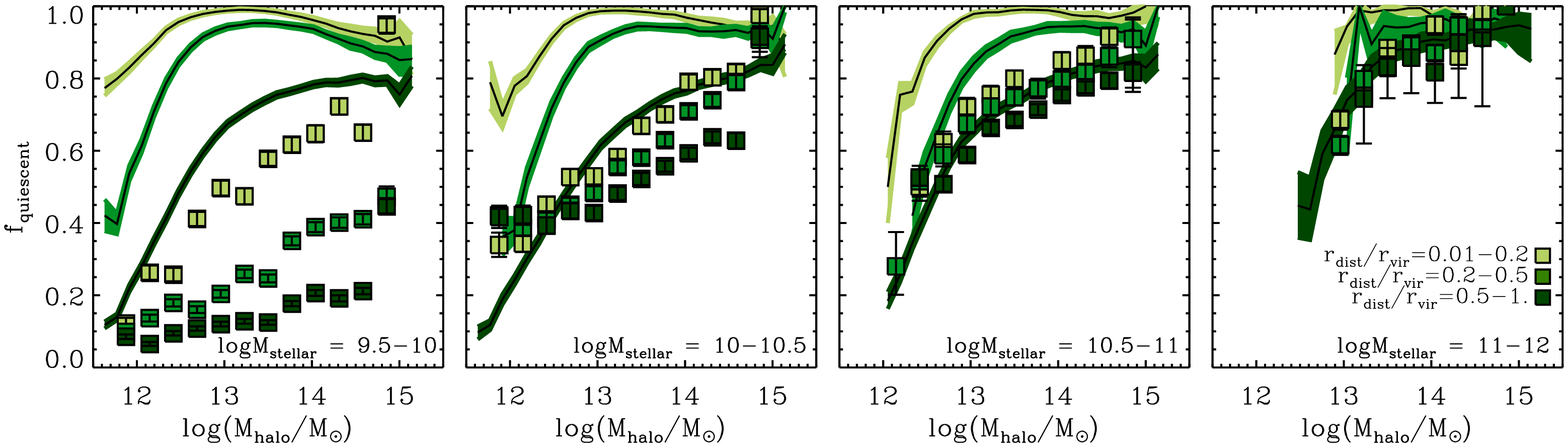,
    width=1.0\textwidth}\vspace{-0.5cm}  
  \caption{Quiescent fractions of satellite galaxies divided into bins 
    of their radial distance to the halo centre (different colours and
    line thickness as indicated in the legend) are plotted versus the
    parent halo mass for observations (symbols, increasing line
    thickness with increasing distance to the halo center) and the Guo
    model (lines with shaded areas). Different panels  correspond to
    different stellar mass bins. Interestingly, observations mainly
    reveal a dependence on the radial distance for lower mass satellites. }  
 {\label{Quiesc_distsat_halo}}
\end{figure*}

As central galaxies with masses $\log (M_{\mathrm{stellar}} / M_\odot)
= 10-11$ typically have halo masses of $\log (M_{\mathrm{halo}} /
M_\odot) = 11.5-12.5$ with virial radii of $\sim 100-200$~kpc,
the observed environmental trends out to 1~Mpc would imply
that effects on super-halo scales should be relevant out to $5-10
\times r_{\mathrm{vir}}$.  
This is in agreement with the study of \citet{Wetzel14} who find using
SDSS data an increased quiescent central fraction out to two times the
virial radius of the host halo (see their Fig. 1): a massive
cluster ($\log (M_{\mathrm{halo}} / M_\odot) = 14$) has typically a
virial radius of $\sim 500$~kpc so that the quiescent fraction is
increased out to a distance of 1~Mpc (from the cluster centre).    
 
In the top row of Fig. \ref{Quiesc_cent_sh} we compare models
  (lines and shaded areas) and observations (symbols, increasing line 
  thickness with decreasing density) for the
  quiescent fraction of centrals versus the small-scale 0.2~Mpc
  density, binned in different large-scale 0.2-1~Mpc annulus densities
  (different colours) and different bins of stellar mass (different
  panels). While observations show the typical, strong residual
  effects for \textit{all} stellar mass bins on top of the 0.2~Mpc
  density, models reveal a clear dependence only at low stellar
  masses. This trend in the model predictions is similar to the
  observed one at low densities, but less pronounced than in the
  observations at high densities. For more massive galaxies ($\log
  (M_{\mathrm{stellar}} / M_\odot) > 10$) models do not predict any
  residual trend -- in contrast with observations. 

In the middle and bottom row of Fig. \ref{Quiesc_cent_sh}, we again
split the central galaxy sample into backsplash and {\it true} centrals,
respectively, to investigate their contribution to the overall
residual effects shown in the top row. As before, the backsplash and
{\it true} central quiescent fraction are estimated with respect to the
total amount of central galaxies (see equation \ref{bspfrac}). This
clearly shows that \textit{any residual density effect on scales above
  0.2~Mpc in the model is caused by the backsplash population.} 

Low-mass, {\it true} central model galaxies do not reveal any residual
effect of the large-scale density on top of a given small-scale
density. For more massive {\it true} centrals, an anti-correlation
between their quiescent fraction and the large-scale density emerges:
at a given small scale density, the fraction of quiescent and {\it
  true} centrals increases with decreasing large-scale density. This
behaviour can be understood as follows: the fraction of quiescent and
{\it true} centrals (given by equation \ref{suppos}) is the
product of the fraction of {\it true} centrals which are quiescent and
the fraction of all centrals which are {\it true}. The former is not
  dependent on the (large-scale) density, while the latter (the {\it
    true} fraction of centrals) increases with decreasing large-scale
  density (see bottom row in Fig.  \ref{Quiesc_cent_bp}). This means
  that the more isolated central galaxies are (lower density), the
  higher is the  {\it true} fraction of centrals and thus, the
  probability that they have always been central galaxies. As a
  consequence, the plotted fraction
  ${n_{\mathrm{qu,true}}}/{n_{\mathrm{cent}}}$ will increase with
  decreasing large-scale density (simply because the {\it true}
  central fraction is increasing).

The discrepancy between observations and models with respect
to the density dependence of quiescent centrals on super-halo scales
between 0.2 - 1~Mpc indicates that models might miss environmental
effects working on centrals. Possible solutions will be discussed in
section \ref{discussion_centrals}.

%**********************************************************************************
\subsection{Satellite galaxies}\label{satellites}
%**********************************************************************************

In this section, we turn to quiescent satellite galaxies and explore the
physical origin \textit{of the density dependence of the quiescent
  fraction of satellites}. We have already pointed out in section
\ref{density_halo} that their density dependence is partly caused by
the parent halo masses which strongly correlate with the 1~Mpc density
in both observations and models. We now examine the role of the
projected radial distance of satellites to their parent halo centre.

%**********************************************************************************
\subsubsection{The relation between density and the projected radial distance to the parent halo centre }
%**********************************************************************************

Fig. \ref{Dens_distsat} shows that the 1~Mpc density above $\log
(\Sigma_{1\mathrm{Mpc}}+1) > 0.5$ strongly correlates with the radial
distance of the satellites to the centre of their parent haloes
(normalised to the virial radius of the parent haloes) in both
observations (symbols) and models (solid lines with green shaded
areas): as expected, satellites closer to the centre of their parent
halo are generally residing in denser environments. The relation
becomes slightly weaker for higher stellar masses (shown by the
different panels). Observations reveal a slightly steeper relation at
galaxy masses below $\log (M_{\mathrm{stellar}} / M_\odot) < 11$. Such
a correlation between density and radial distance to the halo centre
is also in agreement with the study of \citet{Woo13} (see their
Fig. 6) although they use a different density estimator. 

For low-density satellites ($\log (\Sigma_{1\mathrm{Mpc}}+1) < 0.5$,
i.e. less than three neighbours within a 1~Mpc cylinder), the radial
distance is decreasing with decreasing density. This behaviour is
driven by the regime of isolated compact groups, triplets and pairs:
the smaller the number of group members is, the smaller and the more
compact the groups are, reducing the relative radial distance to the
halo centre. 

These results imply that the density dependence of quiescent
satellites is not only driven by the parent halo mass alone, but also
by a trend with radial distance. 
It should be noted that there is also a correlation between radial
distance and accretion time: satellites closer  to the centre were
accreted on average earlier than those residing at larger distances
from the halo centre. Therefore, satellites closer to the centre have
been subject to the environment of the halo for longer time.

%**********************************************************************************
\subsubsection{The dependence of the quiescent satellite fraction on the radial distance }
%**********************************************************************************

Fig. \ref{Quiesc_distsat_halo} shows the quiescent satellite fraction
versus halo mass in different radial distance bins (as indicated in
the legend) for different stellar mass bins (different panels).
Observations reveal a strong residual dependence on the radial
distance for low mass satellites ($\log (M_{\mathrm{stellar}} /
M_\odot) < 10.5$) in parent haloes with masses above $\log
(M_{\mathrm{halo}} / M_\odot) > 13$. This is consistent with the
observational results presented in \citet{Woo13}.   

This may suggest that environmental processes like strangulation,
ram-pressure stripping etc. are most efficiently working on lower mass
satellites, but seem to be of less relevance for more massive
satellites. The latter may instead be more strongly influenced by internal
quenching processes (as stellar or AGN feedback). This agrees roughly
with results by \citet{Peng12} who show that quiescent fractions of
low mass galaxies are mainly dependent on density (and not on stellar
mass) indicating the relevance of environment for star formation
quenching in this regime.

We have verified (not shown here) that at fixed density and stellar
mass, there is hardly any residual dependence on parent halo mass or
on radial distance. This means that the increasing fraction of
quenched satellites as a function of density is fully explained by the
trend as a function of the parent halo mass and radial
distance. Equivalently, the halo mass and radial trends can be viewed
as components of the trend with density.

In contrast to the observations, model satellites show a significant
residual dependence on the radial distance on top of the host halo
mass up to large stellar masses of $\log (M_{\mathrm{stellar}} /
M_\odot) = 11$. In addition, models extremely over-estimate the
quiescent satellite fractions at a given halo mass and radial
distance, particularly in the innermost parts of their parent
haloes. Low-mass model satellites in low-mass host haloes
have a strong dependence on the radial distance, while most of the 
satellites residing in massive host haloes are already quenched
irrespectively of the radial distance. In contrast, observations
reveal just the opposite trends.

Overall, these results confirm and strengthen our earlier conclusion
that model satellites, particularly the low mass ones, suffer from too
strong environmental effects leading to too short quenching
time-scales. Moreover, these strong environmental effects predict
excessive residual dependence on the radial distance for $\log
(M_{\mathrm{stellar}} / M_\odot) > 10$ which is not displayed by
observations.

%**********************************************************************
\section{Constraining quenching time scales for satellites}\label{envhist}
%**********************************************************************

In this section, we estimate which time scales for quenching star formation in
satellite galaxies \textit{should} be predicted by the models so that their
quiescent fractions would be consistent with observations. For that, we take
advantage of our knowledge of the environmental history of model galaxies and
correlate it with observational estimates for the total fraction of
  satellites that became quiescent after being accreted. These fractions will
  be referred to as ``transition fractions'' and are defined in Section 5.2
  below.

To constrain quenching time-scales we make the simple assumption that
  galaxies are quenched after having spent a given amount of time in a halo
  more massive than some critical threshold. Comparing the theoretical
  estimates with the observed transition fractions allows us to constrain both
  the typical time-scale for quenching and the typical environment (halo mass)
  where satellite galaxies get quenched.

The fraction of satellites that became quiescent \textit{only after} the
  infall into their host halo is \textit{not} a directly observable
  quantity. Indeed, the observed quiescent satellite fraction results from a
  \textit{superposition of environmental and internal quenching processes}
  (some of the quiescent satellites have already been quenched whilst
  they are still centrals due to internal processes like e.g. supernovae or AGN
  feedback). Unfortunately, estimating the transition fractions from
  observations is not straightforward and, as we will discuss below, requires a
  number of assumptions. Below, we will discuss in detail the caveats in
  previously adopted definitions of the transition fractions, and will attempt
  to improve these estimates. Albeit improved, our method is still
  approximate. For example, one should account for the fact that also
  \textit{after} infall of a star-forming galaxy, internal processes could be
  responsible for (or contribute significantly to) quenching its star
  formation. We will discuss the limitations of our approach in the following.

%**********************************************************************
\subsection{Quantifying the environmental history - environmental
  fractions}\label{envfrac}  
%**********************************************************************

To quantify the environmental history of model satellites,
we have defined their ``environmental fractions'' as \textit{the
  fractions of satellite galaxies (at a given present-day stellar mass
  and density) which have been residing as a satellite in a halo more
  massive than $X$ for a time longer than $Y$.} For $X$ and $Y$, we
consider the values $\log (M_{\mathrm{halo}} / M_\odot) = 12, 12.7, 13$ and
$3,5,7$~Gyr, respectively.  

Fig. \ref{Time_halo_sat} illustrates such environmental fractions
(green lines) versus their present-day 1~Mpc density for different
stellar mass bins (top row: $\log M_{\mathrm{stellar}} = 9.5-10$;
middle row: $\log M_{\mathrm{stellar}} = 10-10.5$; bottom row: $\log
M_{\mathrm{stellar}} = 10.5-11$).  Different columns indicate
different time limits $Y$, while different line styles correspond to
different halo mass limits $X$, as indicated in the legend.
  
At fixed stellar mass and time ($Y$), the model environmental
  fractions increase with increasing present-day density for each
  value of halo mass $X$. At high density, the difference between lines
  corresponding to different values of $X$ becomes small: a very high
  fraction of galaxies in this density range has spent significant
  amount of their lifetime in massive haloes. At fixed stellar mass,
  density and time ($Y$), model fractions decrease with increasing
  $X$. This is expected as satellites selected using a higher halo
  mass limit are just a sub-sample of those corresponding to a lower
  mass limit. For larger values of $Y$, model fractions always
  decrease for fixed density, stellar mass and value of $X$.

At fixed $X$ and $Y$, the environmental fractions increase with
  increasing stellar mass at low densities. For example, more than
  70~per cent of galaxies with masses $\log (M_{\mathrm{stellar}} /
  M_\odot) = 10-10.5$ have been residing as a satellite in haloes more
  massive than $\log (M_{\mathrm{halo}} / M_\odot) > 12$ for more than
  3~Gyr (solid line in the middle left panel). For galaxies with
  masses $\log (M_{\mathrm{stellar}} / M_\odot) = 9.5-10$ this
  fraction decreases to 50~per cent at low densities (solid line in
  the top left panel of Fig.~\ref{Time_halo_sat}). Otherwise, we
  hardly find any dependence of the environmental fractions on stellar
  mass. 
%
%**********************************************************************
\subsection{Quenching time scales derived from transition fractions a
  la ``vdB''}\label{vdB} 
%**********************************************************************

To obtain an \textit{observational} estimate for the fraction of
  galaxies that have been quenched only after they became satellites,
  we begin by using the same approach adopted by
  \citet{vandenBosch08}. In particular, we compute their ``transition
fractions'' as follows:  
\begin{equation}\label{transfrac}
f_{\mathrm{trans}} = \frac{(f_{\mathrm{sat,qu}}(z_0,M_{z0}) -
  f_{\mathrm{cent,qu}}(z_{\mathrm{inf}},
  M_{\mathrm{inf}}))}{f_{\mathrm{cent,sf}}(z_{\mathrm{inf}},
  M_{\mathrm{inf}})},   
\end{equation}
where $f_{\mathrm{cent,qu/sf}}(z_{\mathrm{inf}}, M_{\mathrm{inf}})$ is
the quiescent (resp. star-forming) central fraction at the time of
infall and for a given stellar mass at infall and
$f_{\mathrm{sat,qu}}(z_0,M_{z0})$ is the quiescent satellite fraction
at present time. The fractions are calculated with respect to the
total number of satellite/central galaxies according to equation
\ref{frac}. The numerator of equation \ref{transfrac} determines the
fraction of all satellite galaxies that have undergone a star-forming
to quiescent transition after their accretion. This decreases with
increasing stellar mass simply reflecting that at the massive end,
most central galaxies lie already on the ``quiescent sequence''. The
denominator determines the fraction of infalling galaxies which were
star forming at the time of accretion (which also depends upon mass)
-- and thus equation \ref{transfrac} gives us the total fraction of
galaxies which have become quiescent since accretion.  Hence,
  the transition fractions give an approximate estimate of the
  fraction of satellite galaxies that have been quenched by
  environmental processes.

\begin{figure*}
  \centering
  \epsfig{file=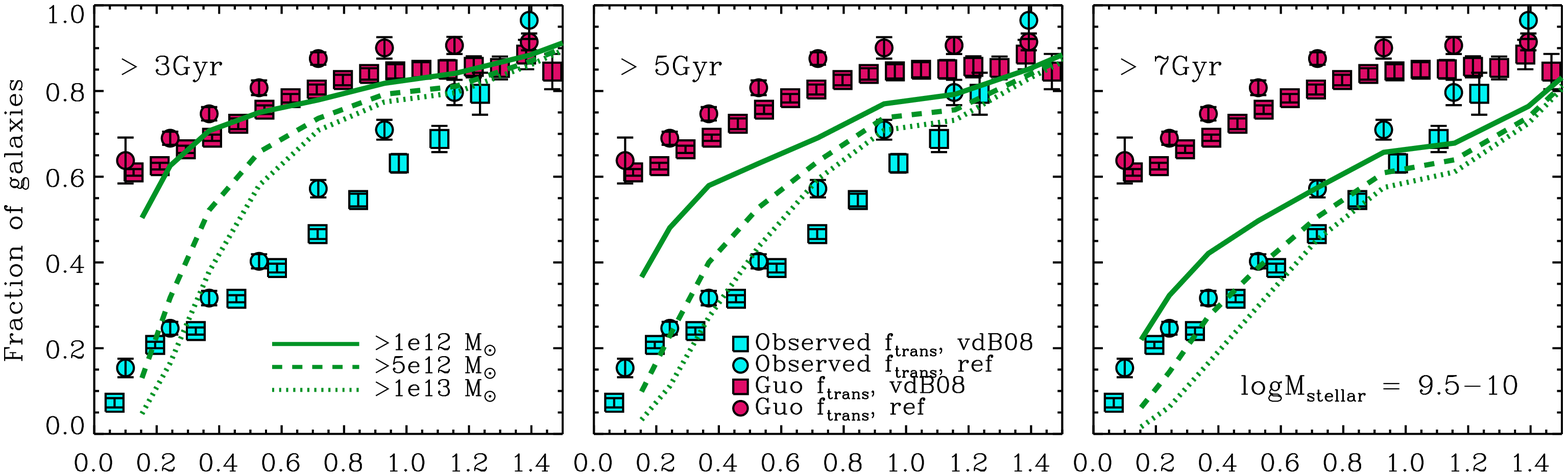,
    width=1.0\textwidth}\vspace{-0.8cm} 
  \epsfig{file=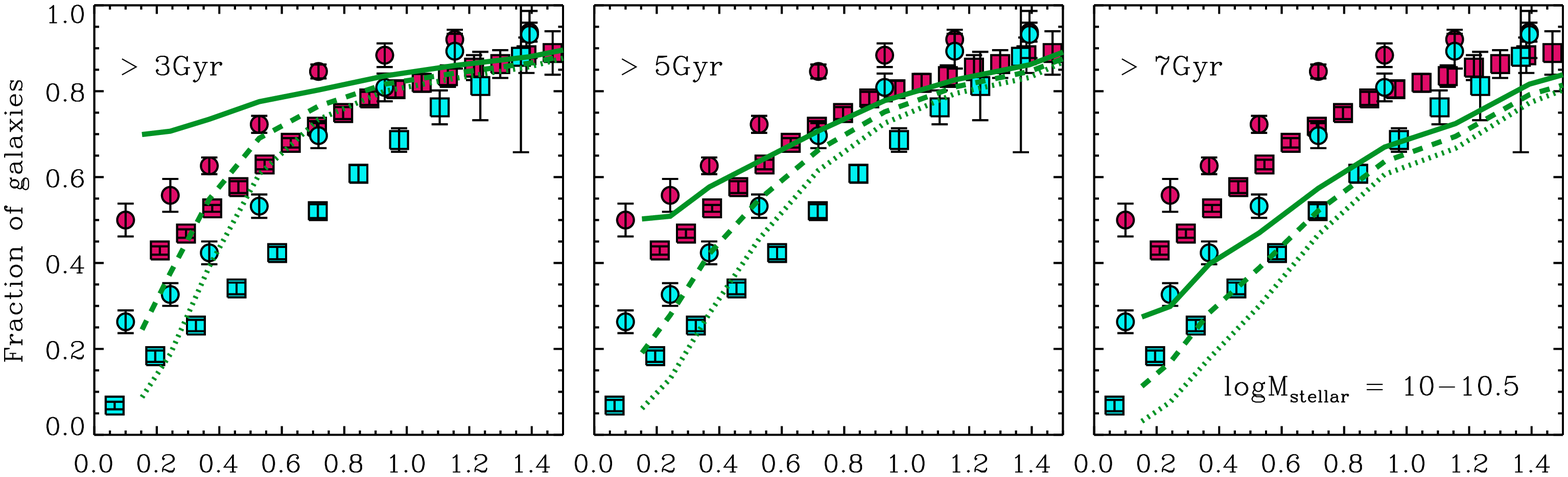,
    width=1.0\textwidth}\vspace{-0.8cm} 
  \epsfig{file=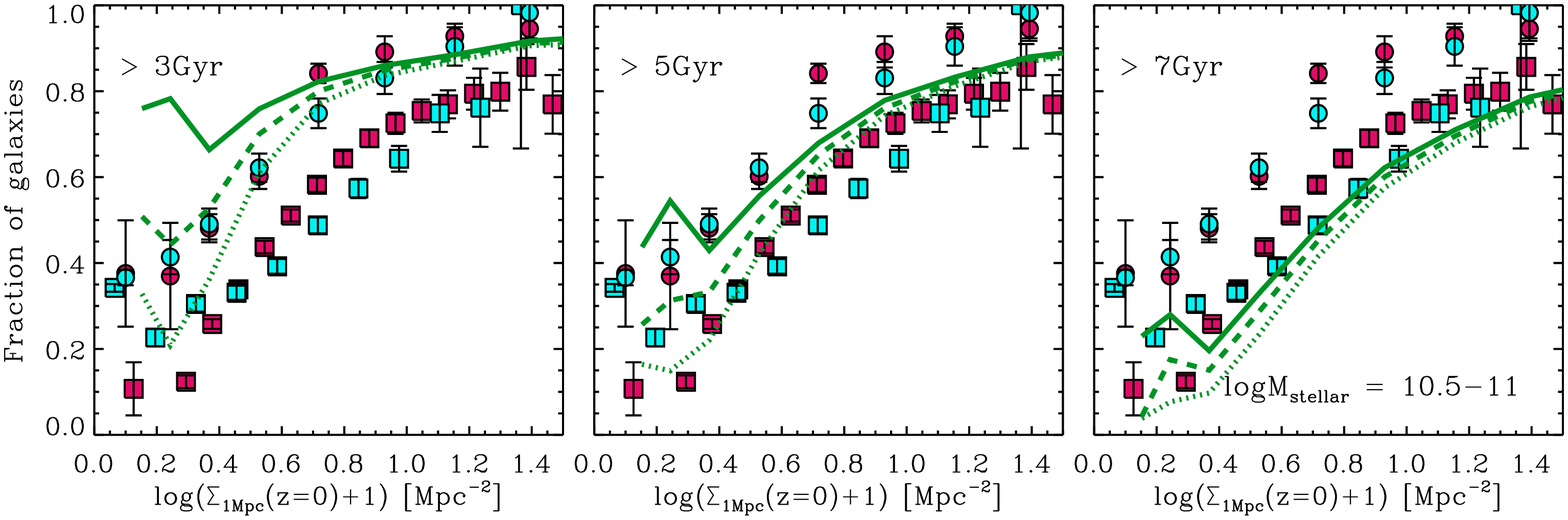,
    width=1.0\textwidth} 
 \caption{Fraction of model satellite galaxies having spent as a
   satellite more time than Y in haloes more massive than X (different
   line styles in green color) versus the present-day 1~Mpc density
   for different stellar mass bins (different rows). Different columns
   correspond to different values of time Y. The squares illustrate
   the "vdB" transition fractions, the circles the refined
   ones. Magenta symbols correspond to the transition fractions
   derived from the Guo model, while the cyan ones illustrate them
   derived from the observational data. Comparing the observed refined
   transition fractions with modelled environmental fractions suggests
   for low mass satellites long quenching  time scales of $\sim$5~Gyr
   which are decreasing with increasing stellar mass.}   
 {\label{Time_halo_sat}}
\end{figure*}

In the study of \citet{vandenBosch08}, they make the simplified
assumption that the quiescent central fraction at the time of infall
is the same as at the present-day:  
\begin{eqnarray}\label{simp}
f_{\mathrm{cent,qu}}(z_{\mathrm{inf}},M_{\mathrm{inf}}) = f_{\mathrm{cent,qu}}(z_{0},M_{0})
\end{eqnarray}
A further caveat in their study is that they assume no strong
  evolution of the stellar mass after infall (within the limits of the
  stellar mass bin considered). These approximations for estimating
  transition fractions have been used in several recent studies
  (e.g. \citealp{Peng12, DeLucia12, Kovac14}). In our study (as we
  need the transition fractions as function of density), we
  additionally assume that central galaxies have been residing in
  the lowest density regions when they were accreted, i.e. we take the
  central fractions at a 1~Mpc density of
  $\log(\Sigma_{1\mathrm{Mpc}}+1) \sim 0.1$.

Comparing now the observed transition fractions with the
  environmental fractions defined in section 5.1, allows us to
  constrain the values of $X$ and $Y$, i.e. the typical environment
  where satellite galaxies are quenched, and the characteristic
  timescale of this process.

The ``vdB'' transition fractions extracted from observations are
illustrated by the cyan squares in Fig. \ref{Time_halo_sat}. They best
match the model environmental fractions (green lines) where galaxies
have been satellites for more than $5-7$ Gyr, almost irrespective of
stellar mass and the parent halo mass in which they have been living
in the past (see first column in table \ref{tab_quenchtime}).  

Galaxies in our stellar mass range which have been satellites for
$>7$~Gyrs almost all live in haloes with $\log (M_{\mathrm{halo}} /
M_\odot) > 13$ (see the similarity of green lines in right panel of
Fig.~\ref{Time_halo_sat}). This implies that the quiescent fraction is
almost entirely determined by the time spent as a satellite in such
haloes. The fraction of satellites in even higher parent halo masses
(e.g. $\log (M_{\mathrm{halo}} / M_\odot) > 14$) declines
because of the paucity of such haloes $\sim 5-7$~Gyr ago.

%**********************************************************************
\subsection{Testing the reliability of the vdB transition fractions
  with models}\label{vdB} 
%**********************************************************************

We now explicitly test the reliability of the approximation used
  in the previous section by taking advantage of the semi-analytic
  model results: first, we treat predictions from the Guo model like
observational data, and estimate typical quenching time scales by
calculating the ``vdB'' transition fractions according to equations
\ref{transfrac} and \ref{simp}. Such model ``vdB'' transition
fractions are illustrated in Fig. \ref{Time_halo_sat} by the magenta
squares. Their best match with environmental fractions suggests
quenching time scales $< 3$~Gyr for satellites with
$\log(M_{\mathrm{stellar}}/M_\odot) = 9.5-10$, $\sim 3-4$~Gyr for
satellites with $\log(M_{\mathrm{stellar}} /M_\odot) = 10-10.5$ and
$\sim 5$~Gyr for satellites with $\log(M_{\mathrm{stellar}}/M_\odot) =
10.5-11$ (see third column in table \ref{tab_quenchtime}).

\begin{table*}
\caption{Summary of the quenching time scales (in Gyr) estimated by the
  different methods (first and second column: vdB and refined
  transition fractions based on observations; third and fourth column:
  vdB and refined transition fractions based on the Guo model; fifth
  column: directly derived quenching time-scales from the Guo model)
  for a given stellar mass bin (in $\log M_{\mathrm{stellar}}$).} 
\begin{tabular}{p{2.5cm} p{2.2cm} p{2.2cm} p{1.6cm} p{1.6cm} p{2.cm}}
\hline\\\bf{Stellar mass bin} & \bf{Observed ftrans vdB08} &
\bf{Observed ftrans ref} & \bf{Guo ftrans vdB08} & \bf{Guo ftrans ref}
& \bf{directly calculated from Guo model}\\  
\hline \hline 
9.5-10.0 & 6 & 5-6 & $<$ 3 & $<$ 3 & 2.5 \\
10.0-10.5 & 6 & 3-5 & 3-4 & 3 & 2.9 \\ 
10.5-11.0 &  7 & 3 & 5 & 3 & 2.9 \\ 
\hline
\end{tabular}
\label{tab_quenchtime}
\end{table*} 

Second, we compute the average quenching timescales of model
  satellites by using the full assembly and merger histories of model
  galaxies. In particular, for each model galaxy, we trace its most
  massive progenitor back in time and compute the time since when the
  galaxy was last a star-forming central until it was for the first
  time a quenched satellite. If the vdB approximation were correct, we
  would, of course, expect to obtain \textit{the same} quenching
  time-scales by the direct and indirect (vdB) method. For low mass
  satellites the directly derived quenching time-scale is in
  rough agreement with the one estimated from the ``vdB'' transition
  fractions ($2.5$~Gyr). For more massive satellites
($\log(M_{\mathrm{stellar}}/M_\odot) = 10-11$), however, models have
satellite quenching time-scales of $2.9$~Gyr (see last column in table
\ref{tab_quenchtime}), significantly shorter than those predicted by
the ``vdB'' approach based on the Guo data (third column in table
\ref{tab_quenchtime}). 

This discrepancy likely has its origin in the simplified
  assumptions at the basis of the ``vdB'' transition fractions. In
  particular, the quiescent central fraction at the time of infall may
  be different to the present-day one. The observational situation is
  not clear, e.g. \citet{Knobel13} do not find any redshift evolution
  of the red central fraction at a given stellar mass, while
  \citet{Tal14} find a strong z-dependence for the cumulative red
  central fractions. To overcome some of the limitations of the
  ``vdB'' transition fractions, we have defined and estimated
``refined'' transition fractions $f_{\mathrm{trans,ref}}$ that we
detail in the following section. We note that part of the
  discrepancy could be also due to the assumed strong correlation
  between the time spent in a halo and the efficiency of quenching
  (i.e. to the very same definition of our environmental fractions).

%**********************************************************************
\subsection{Quenching time scales derived from refined transition
  fractions}\label{ref} 
%**********************************************************************

For the calculation of the ``refined'' transition fractions
$f_{\mathrm{trans,ref}}$ we also use equation \ref{transfrac}, but in
contrast to \citet{vandenBosch08}, \textit{we now explicitly compute
  the quiescent (resp. star-forming) central fraction at the time of
  infall} by following the estimation of a recent study of
\citet{Wetzel13}:     
\begin{eqnarray}
f_{\mathrm{cent,qu}}(z_{\mathrm{inf}}, M_{\mathrm{inf}})
= \frac{f_{\mathrm{all,qu}} -
  f_{\mathrm{sat,qu}}f_{\mathrm{sat}}}{1-f_{\mathrm{sat}}}.
\end{eqnarray}
Here, all quantities are at the time of infall for a given infall
stellar mass, $f_{\mathrm{all,qu}}$ is the quiescent fraction of 
all galaxies ($=n_{\mathrm{qu}}/n_{\mathrm{all}}$) and $f_{\mathrm{sat}}$  
is the satellite fraction ($=n_{\mathrm{sat}}/n_{\mathrm{all}}$). As
observations do not provide all required information, we combine
information on the satellite accretion history and fractions from the
models (which are mainly based on dark matter N-body simulations and
should be, thus, robust and model independent) with the
evolution of the total quiescent fraction in the observations: 
\begin{itemize}
\item  First, for a given present-day stellar mass and
  density we estimate the average infall times and infall masses of
  satellites using the galaxy merger trees from the models and
  following back in time the progenitor galaxies on the main branch.  
\item  Second, we use observed stellar mass functions of
quiescent and star-forming galaxies (\citealp{Ilbert13}) to estimate
the quiescent fractions of all galaxies $f_{\mathrm{all,qu}}$ at the
average time of infall for a given average infall stellar mass
(calculated from the assembly history of the models). 
\item  Third, we extract the satellite fraction $f_{\mathrm{sat}}$ at 
  the time of infall (and a given stellar mass bin at that time) from
  the models.  
\end{itemize}

\begin{figure*}
  \centering
  \epsfig{file=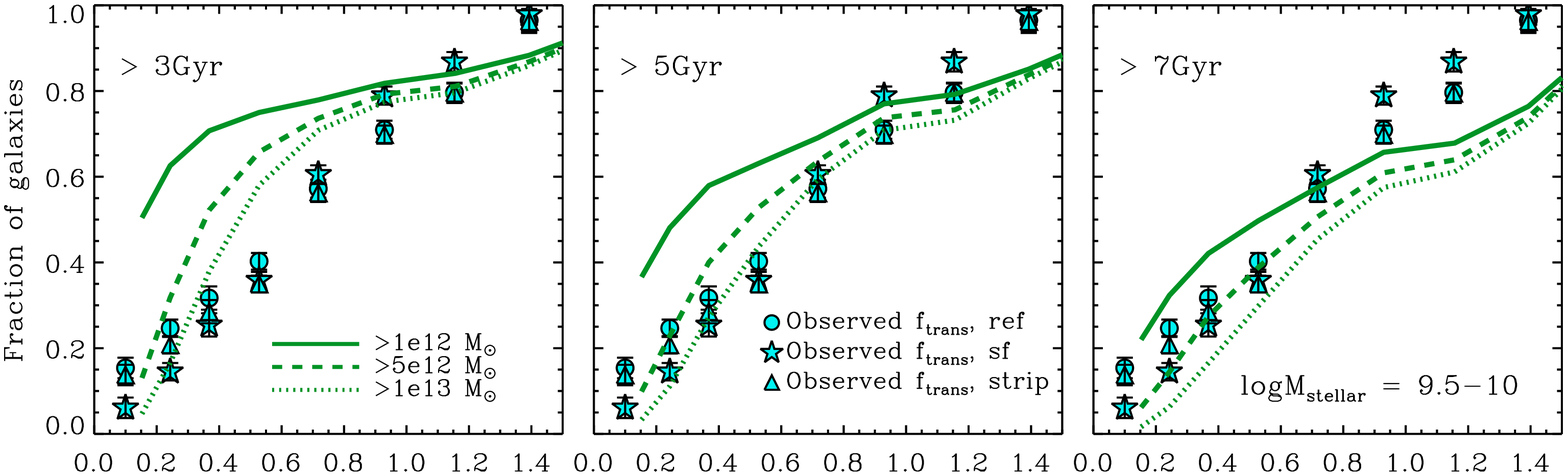,
    width=1.0\textwidth}\vspace{-0.8cm} 
  \epsfig{file=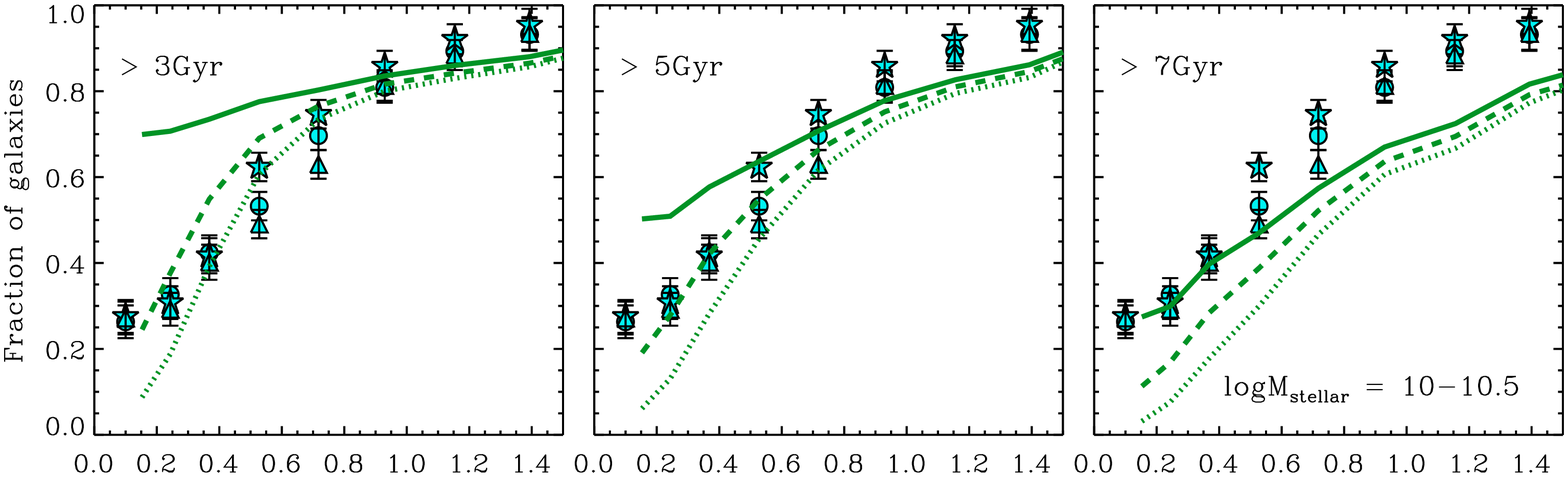,
    width=1.0\textwidth}\vspace{-0.8cm} 
  \epsfig{file=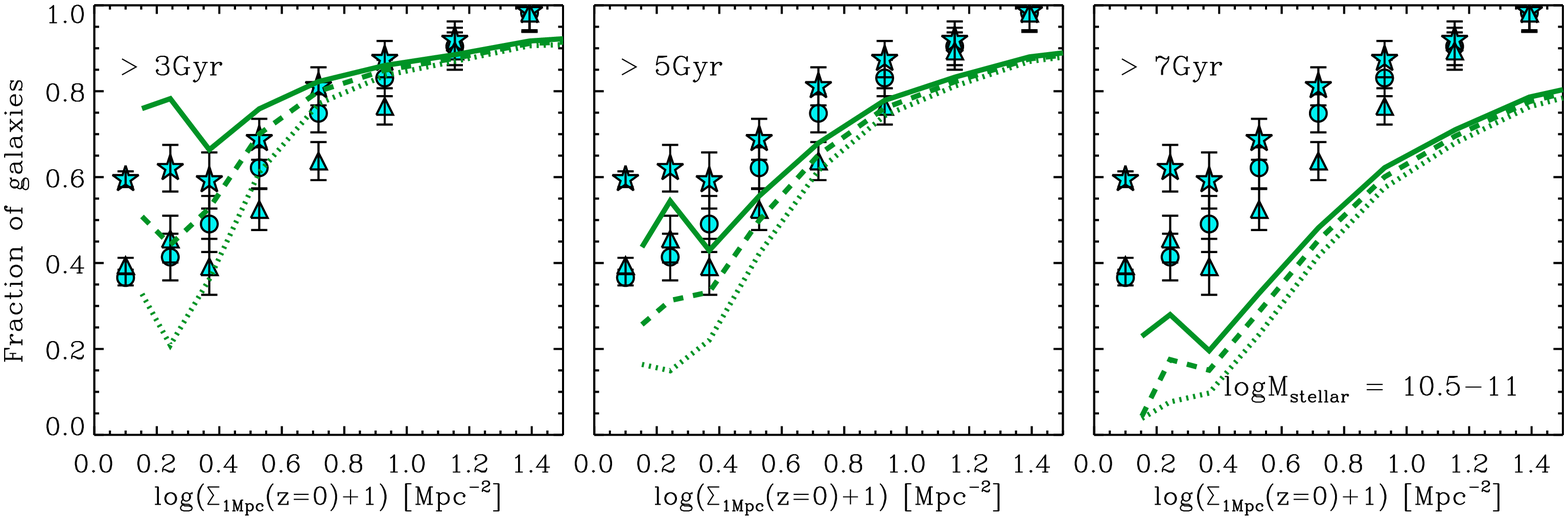,
    width=1.0\textwidth} 
 \caption{Same as in Fig. \ref{Time_halo_sat}, but in addition, the
   cyan stars and triangles illustrate the refined transition
   fractions derived from observations when two extreme scenarios for
   the evolution of the stellar mass after infall are assumed: either
   the satellite stellar mass decreases due to stellar stripping or it
   increases due to continued star formation. The inaccuracy arising
   from these processes does not influence the resulting
   quenching time scales for satellites.}  
 {\label{Time_halo_sat_test}}
\end{figure*}

In addition, we impose no evolution of the quiescent satellite
fraction $f_{\mathrm{sat,qu}}$, i.e. we assume that it is the same at
the time of infall as in today's Universe:
\begin{equation}
f_{\mathrm{sat,quench}}(z>0) = f_{\mathrm{sat,quench}}(z=0)
\end{equation}
 This choice is justified by \citet{Tinker10} who found no evolution
 in the quiescent fraction for satellites at fixed magnitude at $z
 \leq 1$ based on halo occupation modelling. We have also explicitly
 tested the effect of an evolving quiescent satellite 
 fraction with redshift using the results of the observational studies
 of \citet{Mok13} and \citet{McGee11}, and find that the results do
 not change significantly.

The ``refined'' transition fractions are illustrated by the cyan
circles in Fig. \ref{Time_halo_sat}. For low mass satellites $\log
(M_{\mathrm{stellar}} / M_\odot) < 10$, they are in agreement with the
``vdB'' transition fractions and suggest quenching times of $\sim
5$~Gyr. For more massive satellites, the differences between the
``vdB'' and the ``refined'' transition fractions become larger, and
the latter ones suggest quenching time scales of $3-5$~Gyr for
satellites with $\log (M_{\mathrm{stellar}} / M_\odot) = 10-10.5$ and
relatively short quenching time scales of $\sim 3$~Gyr for massive
satellites with $\log (M_{\mathrm{stellar}} / M_\odot) =
10.5-11$ (see second column in table \ref{tab_quenchtime}). Therefore,
the refined fractions suggest quenching time-scales that depend on
stellar mass (are longer for low mass satellites than for massive ones).

To again test the reliability of the refined approach, we compute - as
before for the ``vdB'' method - the refined transition fractions of
the Guo model (see magenta circles in Fig. \ref{Time_halo_sat}). They 
suggest quenching time scales of $< 3$~Gyr for satellites with
$\log(M_{\mathrm{stellar}}/M_\odot) = 9.5-10$ and of $\sim 3$~Gyr for
satellites with $\log(M_{\mathrm{stellar}}/M_\odot) = 10-11$ (see
fourth column in table \ref{tab_quenchtime}). These estimates are now
consistent with the quenching time scales directly derived from the
Guo model demonstrating \textit{the improvement of this refined
  method}.  

We also point out that for massive
satellites with $\log(M_{\mathrm{gal}}/M_\odot) = 10.5-11$ the refined
transition fraction from the Guo model and from the observations are
in reasonably good agreement which reflects the fact that the models
match the observed density dependence of the quiescent satellites in
that mass range (see top row in Fig. \ref{Quiesc_dens_Guo}).    

%**********************************************************************
\subsection{Caveats}\label{caveats}
%**********************************************************************

We have demonstrated that the refined approach for constraining
quenching time scales is an improvement compared the ``vdB'' method,
but there is still one caveat: how the stellar mass of a satellite
galaxy evolves after infall (and thus, in which stellar mass bin
satellites end at $z=0$) is strongly model dependent. Satellites can
experience stellar stripping (e.g. \citealp{Pasquali10}) and
therefore, be less massive at $z=0$, or/and can continue forming stars
at a significant rate and therefore, become more massive at present.

The Guo model includes some prescriptions for both processes, but the
extent to which their predictions are robust is unclear. To assess
the influence of uncertainties of these prescriptions, we consider two
extreme scenarios: first, only a decreasing stellar mass due to
stellar stripping and second, only an increasing stellar mass due to
residual star formation. In these scenarios, the same satellite galaxy
can end up in a different stellar mass bin at $z=0$ which may
influence the final estimation of the quenching time scales.

For the first scenario, we compute the average decrease in stellar
mass (due to stellar stripping) as a function of the stellar mass at
time of infall by using a fitting formula extracted from controlled
simulations of galaxy groups (\citealp{Contini14};
Villalobos et al., in prep.):  
\begin{eqnarray}
M_{\mathrm{strip}}(t) = M_{\mathrm{st}}(t_{\mathrm{inf}})
\exp \left[ \frac{-16}{1-\eta}
  \sqrt{\frac{M_{\mathrm{sub}}}{M_{\mathrm{par}}}} \left( 1 -
        \frac{t}{t_{\mathrm{merg}}} \right) \right]
\end{eqnarray}
Here, $M_{\mathrm{st}}(t_{\mathrm{inf}})$ is the stellar mass at the time
of infall, $\eta$ is the circularity of the orbit, $M_{\mathrm{sub}}$
and $M_{\mathrm{par}}$ are the subhalo and the parent dark matter halo
mass, respectively, and $t_{\mathrm{merg}}$ is the residual merger
time of the satellite galaxies. Merger times are estimated from a
formula (again derived from controlled group simulations) of
\citet{Villalobos13} which is a refinement of the one presented in 
\citet{Boylan08}. As value for the circularity, we use the one of the
most likely infalling orbit ($\eta = 0.51$). The stellar and halo
masses are taken directly from the model at the time of infall. 

For the second scenario (only increasing stellar mass due to residual
star formation), we assign to each infalling galaxy a star formation
rate using the observed relation between the SFR and stellar mass
(\citealp{Elbaz07,Daddi07}) at the time of infall. By assuming that
the star formation rates stay constant after accretion we compute 
the average increase in stellar mass for satellite galaxies since
they have become a satellite. 

Fig. \ref{Time_halo_sat_test} shows the refined transition fractions
(cyan circles) and the ones for the purely star forming and the purely
stellar stripping scenario (cyan stars and cyan triangles,
respectively). For the lowest stellar mass bin (top row), hardly any
differences arise between the transition fractions derived from the
different scenarios. Towards more massive satellites (middle and
bottom row in Fig. \ref{Time_halo_sat_test}), however, the differences
become larger, particularly for satellites with
$\log(M_{\mathrm{stellar}}/M_\odot) = 10.5-11$ residing in low density
regions (up to 50~ per cent larger transition fractions for the
star-forming scenario). 

Nevertheless, the uncertainties in deriving the transition fractions
based on scenarios of a pure increase or a pure decrease in the
stellar mass of an accreted galaxy are not large enough to
change the main result: the resulting quenching time scales when
comparing the different transition fractions to the environmental
fractions are hardly influenced by the stellar evolution of the
satellite after infall.

\section{Discussion and Conclusions}\label{summary}

\subsection{Comparison of our observational results to previous
  studies}

Our observational results for the density dependence of quiescent
satellite galaxies agree qualitatively with previous observational
studies of the present-day Universe (e.g. \citealp{Peng12, Kovac14}
and \citealp{Woo13}). These studies also highlight - even if using different
density estimators - a strong density dependence of the quiescent
satellite fractions at a given stellar mass. 

In contrast, for the density dependence of quiescent central galaxies,
  the situation is less clear: \citet{Peng12} and \citet{Kovac14} find no
  significant density dependence on top of the stellar mass dependence, while
  the quiescent fractions of low mass centrals depend, albeit weakly, on
  density in the study of \citet{Woo13}. These differences may be ascribed to
  different estimators for quiescence (e.g. colour is used in
  \citealp{Peng12}), to different definitions of environment, to redshift
  measurement errors, and to the used tracer/neighbour galaxies and their
  sampling rate (e.g. \citealp{Kovac14} has a significantly lower sampling rate
  than ours). Therefore, it is generally difficult to directly compare our
  observational results with previous ones in the literature. 

Nevertheless, Fig. 1 by \citet{Peng12} shows that they also \textit{do
    find} a weak density dependence of central galaxies at high densities (the
  trend is not as steep as the one for satellites). The statistical
  significance is, however, low because of the low numbers of central galaxies
  at high densities. As noted above, \citet{Peng12} have used optical colours
  to distinguish between quiescent and star forming galaxies, which is less
  accurate than using sSFR (see e.g. \citealt{Woo13}). In addition, they might
  select bluer neighbours (due the B-band observations) which are less
  clustered than redder ones.

In agreement with previous studies, we find that the global
density dependence is mainly driven by satellite galaxies, as the vast 
majority of galaxies residing in denser regions are satellite
galaxies, while the majority ($\sim$85~per cent) of central galaxies
are located in lower density regions. Even if the central galaxies
residing in denser regions show a similarly steep dependence on
environment as satellites, they constitute only a minor contribution
to the overall galaxy population in these regions (but still
non-negligible to the central population). \textit{Therefore, one
  striking result compared to previous studies is the extremely
  similar behaviour of quiescent centrals and satellites at a given
  density and stellar mass.}    

\subsection{Model limitations and future perspectives}

\subsubsection{Satellites}

The over-production of quiescent satellite galaxies in models is a
well-known problem (e.g. shown in \citealp{Kimm09} using different
semi-analytic models) and often attributed to simplified recipes for
environmental processes, such as the assumption of an instantaneous
strangulation.
Our results demonstrate that even the more relaxed, delayed
strangulation assumption in the Guo model is not sufficient to solve
the ``over-quenching problem'' of satellite galaxies with masses below
$\log (M_{\mathrm{stellar}} / M_\odot) < 10.5$.  We have
  verified that a model assuming an instantaneous stripping of the hot
  gas and adopting a different definition for satellite galaxies,
  where environmental processes effectively start when a galaxy enters
  a FOF group, (we have considered the public catalogues based on the
  \citealp{DeLucia07} model) predicts an even larger fraction of
  quiescent satellites up to stellar masses of $\log
  (M_{\mathrm{stellar}} / M_\odot) = 11$. Therefore, the gradual
  stripping (and different satellite definition) assumed in the Guo
  model does influence the evolution of satellite galaxies bringing
  model predictions in better agreement with data. But the effect does
  not appear to be sufficient to solve the problem of the excess of
  passive satellites.

There have also been attempts to improve the modelling of
environmental processes, e.g. by \citet{Font08} or \citet{Weinmann10},
but none of them are completely successful in matching the
observational data. This shows, that \textit{the recipes for
  environmental effects working on satellite galaxies need to be
  further refined}. 

In a recent study, \citet{Henriques13} have shown that a better
  agreement with data can be obtained assuming a different scaling for
  the re-accretion time scale of gas. In their model, low mass
  galaxies form later, are typically younger and more star forming
  than in the Guo et al. model. It will be an important further test
  for this model} to investigate how well it can predict
the global environmental dependencies of quiescent satellites. It is
likely that additional changes for the tidal and ram pressure
stripping will be necessary in order to capture the observed strong
dependence on density for low-mass satellites and predict long
quenching time scales of 5~Gyr for low-mass satellites as estimated in
this study.

\subsubsection{Centrals}\label{discussion_centrals}

We have shown that models cannot predict the similar environmental
behaviour of centrals and satellites in the sense that models cannot
capture the observed strong density dependence of quiescent central
galaxies, particularly the one on super-halo scales
($0.2-1$~Mpc). Two effects should be considered in this case.

On the one hand, \textit{true centrals} might suffer from
environmental effects by directly interacting with an extended hot
halo beyond the virial radius. This effect was e.g. investigated
by \citet{Bahe13} using hydrodynamical simulations and was shown to
lead to a slow stripping of the hot halo component of the infalling
central galaxy. The corresponding effect on the cold gas and star
formation is, however, rather weak in the simulations. Such an
additional effect might help models to capture a residual
environmental effect on $0.2-1$~Mpc scales for galaxies more massive
than $\log (M_{\mathrm{stellar}} / M_\odot) > 10.5$, where {\it true}
centrals are the dominant contribution to the overall central
population.   

On the other hand, backsplash centrals may still feel the
gravitational potential of their parent halo and may, thus, continue 
to experience strangulation, i.e. a stripping of their hot gas, even
if they have left their parent halo again. Or another, maybe
more likely possibility is that backsplash centrals which have passed
through the main (host) halo, no longer live inside a filament which
would provide them with a continuous supply of gas (as with {\it true}
centrals). Therefore, their star formation continues to exhaust their
existing gas content as if they were still satellites.

A recent study of \citet{Wetzel14}, using empirical models
for galaxy formation, has shown that such a continued treatment of
backsplash centrals as satellites results in a good agreement with
observational data. In our models, such an assumption could
essentially increase the density dependence on super-halo scales for
low-mass galaxies and galaxies residing at high densities where the
fraction of backsplash centrals dominates.  Overall, such processes
are not taken into account in the models, but seem to be important to
capture the observed environmental trends on super-halo scales.

%**********************************************************************
\subsection{Satellite quenching time scales and their stellar mass dependence}\label{mass_dep}
%**********************************************************************

\begin{figure}
  \centering
  \epsfig{file=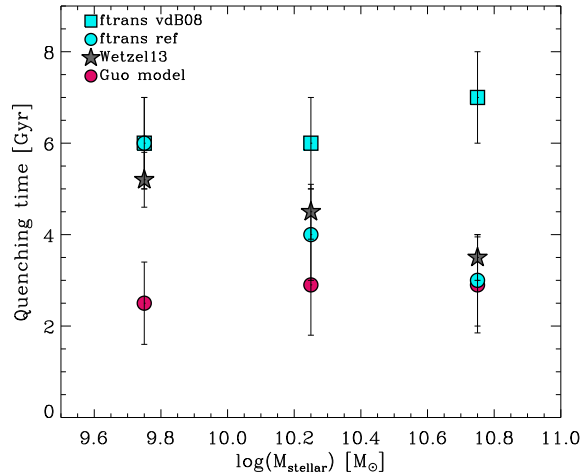, width=0.45\textwidth}
  \caption{Quenching timescales estimated from transition fractions
    ("vdB" method: blue squares, refined method: blue circles) and
    directly from the Guo model (magenta circles) versus stellar masses.
    The quenching times scales estimated from the refined transition
  fractions are in rough agreement with the results from
  \citet{Wetzel13} (grey stars) and are increasing with decreasing
  stellar mass.}  
 {\label{Tquenc_mstellar}}
\end{figure}

Fig. \ref{Tquenc_mstellar} summarises the different
estimates/predictions for the satellite quenching time scales from
section \ref{envhist} summarised in table \ref{tab_quenchtime} showing
them as a function of stellar mass. The ``vdB'' transition fractions
(cyan squares) suggest large quenching time scales of roughly $\sim
5-7$~Gyr independent of mass in agreement with a recent study of
\citet{DeLucia12}.  
However, the ``vdB'' estimate refers simply to the difference between
centrals and satellites at $z=0$, whereas our refined method estimates
the total fraction of satellites {\it quenched since infall}.
The refined transition fractions (cyan circles in
Fig. \ref{Tquenc_mstellar}) -- even when considering the uncertainty
due to star formation and stellar stripping of the satellite stellar
mass after infall -- point towards increasing quenching time scales
with decreasing stellar mass and predict long quenching time scales of
$\sim 5$~Gyrs for low mass satellites.  

In the Guo model (magenta circles in Fig. \ref{Tquenc_mstellar}), the
satellites have typically short quenching time scales, also nearly
independent of stellar mass. They agree with our estimates based on
the SDSS data for massive satellites, but are significantly
under-estimated for low-mass satellites ($\sim 2.5$~Gyr).

Interestingly, quenching time scales predicted by the refined approach
and their mass dependence are in good agreement with the results of a
recent study of \citet{Wetzel13} using a completely different approach
based on an empirical HOD model (see grey stars in
Fig. \ref{Tquenc_mstellar}). In their model, they find a
``delayed-then-rapid'' quenching scenario to reproduce the observed
quiescent satellite fraction and the bi-modality of the SFRs. In their
proposed scenario, satellite SFRs evolve unaffected for the first
several Gyr after infall, after which the star formation quenches
rapidly (assuming an exponential decline).   

Note that also the results of other studies \citet{Skibba09, Kang08,
  Font08, vandenBosch08} imply that satellite quenching processes must
be gradual or delayed and/or that the mechanism (e.g. strangulation)
must operate in both group and cluster-mass haloes.

The stellar mass dependence of the quenching time scales appears
counter-intuitive at first sight as environmental processes 
are expected to more easily affect low-mass satellites due to
their low internal binding energies. The physical origin of this mass
dependence is likely twofold.  

First, internal quenching processes such as AGN feedback are working
more efficiently on more massive satellites than on less massive ones,
reducing their quenching time scales after infall in addition to the
environmental effects they are experiencing. In other words, for more
massive satellites, environmental quenching is not the main/only
quenching effect. This means that the galaxy status does not matter
for massive galaxies to first order. Such an (additional) internally
quenched population would also be consistent with the independence of
the quiescent fractions for massive satellites on the radial distance
as seen in Fig. \ref{Quiesc_distsat_halo}.  The difference between the
``vdB'' or ``refined'' transition fractions is not necessarily enough
to separate such additional internal processes \textit{after
  accretion}. In the (perhaps extreme) case that satellites would
experience the same amount of internal quenching as central galaxies
(at fixed stellar mass), the``vdB'' transition fractions (showing no
stellar mass dependence) would represent the fraction of {\it
  environmentally quenched} satellites as opposed to the {\it total}
fraction of quenched satellites as expressed by the refined transition
fractions. In reality we may expect the fraction quenched due only to
environmental processes to be somewhere between the ``vdB'' and the
refined transition fractions: satellite galaxies with their own sub-halo
probably still experience some (possibly reduced) quenching related to
reduced gas accretion and increased heating from massive galaxies
(i.e. internal processes), similar to central galaxies.

Second, different orbits and dynamical friction time scales for
satellites of different masses may also influence the quenching time
scales. The dynamical friction time scales are strongly dependent on
the ratio between satellite and host halo mass (e.g.
\citealp{Boylan08, Villalobos13}), i.e. the smaller the satellite 
galaxy (at a given host halo mass), the longer is its dynamical
friction time scale. This implies a longer spiralling around in the
outer, low-density regions for low-mass satellites before they get into
the denser, inner regions where environmental processes become more
efficient. The rate at which gas is removed from a satellite is most
likely related to how often it has crossed the higher density medium
along its orbit. This means that a satellite could retain more gas if
it is exposed less often to the high density of the central regions of
the host halo. For the stellar component, this is shown by Villalobos
et al. (in prep.) and it is plausible to assume that the gas will
behave similarly. 

This would also provide a physical explanation for the
``delayed-then-rapid'' quenching scenario of the study of
\citet{Wetzel13}, where the rapid (exponentially decreasing) quenching
may only start when the satellite has reached the dense inner regions
of the halo. In contrast, massive satellites have short dynamical
friction time scales and reach the central, dense regions rapidly so
that their gas gets stripped efficiently in a short time (even
  though their internal potential well is deeper). In addition,
massive galaxies will more rapidly merge with the central galaxy and
then drop out of the satellite population. These will be the ones
which have been in the parent halo longest and so have also been most
likely quenched, but are lost by merging. This means that massive
galaxies cannot have long quenching times because otherwise we would
not see any difference between the massive central (infall) and
massive satellite quiescent fractions as there would be no time for
environmental quenching before they merge away.

The relative contribution of the two scenarios (internal quenching and
orbits/dynamical friction time scales) on the mass dependence of
quenching satellites is unclear. We plan to address this issue in a
future work. 

%**********************************************************************
\subsection{Satellite quenching time scales at higher redshifts}\label{mass_dep}
%**********************************************************************

The analysis in this paper only focuses on the present-day Universe.
At higher redshifts, however, the Universe is apparently not old enough for
such long quenching timescales as we infer at low redshifts. One
explanation for this apparent disagreement is that the currently used
models do not properly account for the internal quenching
processes. Another, possibly complimentary option is that it is very
likely that the quenching timescale scales vary with the dynamical
time, i.e. depend (in addition to halo mass, galaxy mass and density)
on redshift and are thus, shorter at higher redshift. The latter
possibility is in agreement with results from the work by Mok et
al. (submitted to MNRAS) who provide constraints for quenching
time-scales using observations of groups at $z \sim 0.4$ and $z \sim
0.9$. They find that the observed fractions are best matched with a 
model that includes a delay (when quenching starts) that is
proportional to the dynamical time followed by a rapid quenching time
scale of $\sim 0.25$~Gyr.   

%**********************************************************************
\subsection{Summary}
%**********************************************************************

In this study, we have performed a detailed analysis of the effect of
environment and the environmental history on quenching star formation
in satellite and central galaxies. We have taken advantage of publicly
available galaxy formation catalogues where the galaxy formation model
of \citet{Guo11} has been applied to the dark matter Millennium
simulation. The Guo model adopts a gradual strangulation of the hot
gas content of a satellite galaxy, i.e. the hot halo gas is assumed to
be stripped at the same rate as the dark matter halo. We have compared
model predictions to observational data of the present-day Universe
making use of a refined density catalogue of \citet{Wilman10} based on
the SDSS-DR8 database and cross-correlated with the catalogues of
Brinchmann et al. and Yang et al. To select quiescent galaxies we have
used specific SFRs as a tracer, and as an estimator for environment we
have considered the theoretical halo mass and projected densities
within cylinders with two different radii (0.2~Mpc and 1~Mpc).  Our
main results are: 
\begin{enumerate}
\item[{\bf 1.}] \noindent \textit{Observations reveal a surprisingly
    similar behaviour for the environmental dependence of the
    quiescent fraction of centrals and satellites}. Models cannot
  capture the observed trends and predict a significantly different
  behaviour for centrals and for satellites. In particular, they
  significantly over-estimate the quiescent fractions of satellites
  and slightly under-estimate the ones of centrals at fixed stellar
  mass. In addition, models predict a much weaker dependence on
  environment for low-mass galaxies than observed.
\item[{\bf 2.}] \noindent The dependence of (low mass) model and
  observed central galaxies on the 1~Mpc density is not driven by their
  halo masses, but has its origin in environmental effects on
  super-halo scales up to 1~Mpc. The density dependence in the models
  is solely caused by a population of backsplash centrals, which have
  been satellites in the past and thus, could have experienced
  environmental effects over time scales of typically 1.5-3~Gyr. Such
  backsplash centrals preferentially have low masses and reside in
  dense environments.  
  However, observations indicate a stronger density dependence of the
  quiescent centrals on super-halo scales between 0.2-1~Mpc. This
  suggests that models might miss some processes (e.g. a
  ``satellite-like'' treatment of backsplash centrals).  
\item[{\bf 3.}] \noindent The strong density dependence of (low-mass)
  satellite galaxies can be understood as the result of a
  superposition of their host halo mass and their distance to the host
  halo centre, in agreement with e.g. \citet{Woo13}. At a given radial
  distance, host halo mass and satellite stellar mass, models
  significantly over-estimate the quiescent satellite fractions
  compared to observations, indicating that environmental effects are
  working too efficiently on satellites, particularly the low-mass
  ones, resulting in too short gas consumption time scales. 
\item[{\bf 4.}] \noindent Considering the environmental history of model
  satellite galaxies and comparing it to transition fractions derived
  from observational data, we have constrained typical quenching
  time-scales for satellites. We have improved the estimation of
  transition fractions compared to a previous study of
  \citet{vandenBosch08} by explicitly accounting for a time evolution 
  of the quiescent/star-forming fraction of central galaxies. This
  allows for more robust and reliable predictions for quenching time
  scales: we find for low mass satellites long quenching
  time-scales of $\sim 5$~Gyr. Quenching time-scales are
  decreasing with increasing stellar mass. For higher mass galaxies,
  it is less clear, but the quenching timescales seem to be between
  $3-5$~Gyr depending on the method used.
\end{enumerate}

In this paper, we have pointed out some typical failures of currently
used galaxy formation models and with the help of observations we have 
indicated possible directions towards which models should be improved
in future. Our results indicate that models require an improved
prescription of environmental processes and internal processes
\textit{where centrals and satellites should be treated in a much more
  similar way than done so far}: backsplash central galaxies should not
experience further filamentary accretion, while satellite galaxies
which have retained some of their own halo should still experience
AGN/radio-mode/stellar feedback which can lead to internal suppression
of star formation. Within the framework of our models, however, it
might be difficult to obtain long quenching time scales \textit{just}
as  consequence to an improved prescription of environmental
processes. Fundamental changes in the models for star formation,
stellar feedback and re-accretion are likely required.

In addition, our results highlight the highly important role of orbits
and dynamical friction time scales for the quenching time scales of
satellites and backsplash centrals. They may also help to resolve the
apparent contradiction our results imply for centrals and satellites:
while backsplash centrals are supposed to experience stronger
environmental effects to get quenched more efficiently, satellites
should have much longer quenching time-scales resulting in a much
smaller quiescent fraction. Backsplash centrals are sitting on highly
eccentric and thus, radial orbits. They can, therefore, rapidly
encounter one or more peri-centre passages which causes a fast
(continuous) stripping of the gas content. Instead, low mass
satellites on tangential orbits and long dynamical friction time
scales reside for a long time in the outer, low-density parts of their
host halo without any significant stripping of their gas content and
thus, resulting in long quenching time scales and (relatively)
low quiescent fractions.

In future studies, we plan to quantify and to parameterise
the evolution of the gas content of centrals and satellites as a
function of stellar mass, parent halo mass, their orbits and dynamical
friction time scales using controlled simulations of galaxy groups and
clusters. Such parameterisations can be easily implemented in galaxy
formation models, and tested against observations.

\section*{Acknowledgements}

MH, GDL and \'AV acknowledge financial support from the European Research 
Council under the European Community's Seventh Framework Programme 
(FP7/2007-2013)/ERC grant agreement n. 202781. MH additionally
acknowledges support from the European Research Council via an
Advanced Grant under grant agreement no. 321323—NEOGAL. DW
acknowledges support from the Max Planck Gesellschaft (MPG) and the
Deutschen Forschungsgemeinschaft  (DFG). SW acknowledges funding from
ERC grant HIGHZ no. 227749. SZ has been supported by the EU Marie
Curie Integration Grant "SteMaGE" Nr. PCIG12-GA-2012-326466  (Call
Identifier: FP7-PEOPLE-2012 CIG). We thank Xiaohu Yang for providing the
(unpublished) DR7 version of his group catalogues and the
corresponding mock group catalgoue based on the Guo model and we are
grateful to Anna Pasquali and Pierluigi Monaco for fruitful
discussions. Finally, we very much appreciate the careful and
constructive reading of our paper by the anonymous referee.

\bibliographystyle{mn2e}
\bibliography{Literaturdatenbank}

\label{lastpage}

\end{document}